\documentclass[12pt]{article}
\usepackage{amssymb}
\usepackage{graphicx}
\begin{document}
\newcommand{\beq}{\begin{equation}}
\newcommand{\eeq}{\end{equation}}
\newcommand{\beqa}{\begin{eqnarray}}
\newcommand{\eeqa}{\end{eqnarray}}
\newcommand{\beqar}{\begin{eqnarray*}}
\newcommand{\eeqar}{\end{eqnarray*}}
\newcommand{\al}{\alpha}
\newcommand{\be}{\beta}
\newcommand{\del}{\delta}
\newcommand{\D}{\Delta}
\newcommand{\eps}{\epsilon}
\newcommand{\ga}{\gamma}
\newcommand{\Ga}{\Gamma}
\newcommand{\ka}{\kappa}
\newcommand{\nn}{\nonumber}
\newcommand{\inn}{\!\cdot\!}
\newcommand{\h}{\eta}
\newcommand{\ii}{\iota}
\newcommand{\kk}{\varphi}
\newcommand\F{{}_3F_2}
\newcommand{\la}{\lambda}
\newcommand{\La}{\Lambda}
\newcommand{\na}{\prt}
\newcommand{\Om}{\Omega}
\newcommand{\om}{\omega}
\newcommand{\p}{\Phi}
\newcommand{\sig}{\sigma}
\renewcommand{\t}{\theta}
\newcommand{\z}{\zeta}
\newcommand{\ssc}{\scriptscriptstyle}
\newcommand{\eg}{{\it e.g.,}\ }
\newcommand{\ie}{{\it i.e.,}\ }
\newcommand{\labell}[1]{\label{#1}} 
\newcommand{\reef}[1]{(\ref{#1})}
\newcommand\prt{\partial}
\newcommand\veps{\varepsilon}
\newcommand{\pol}{\varepsilon}
\newcommand\vp{\varphi}
\newcommand\ls{\ell_s}
\newcommand\cF{{\cal F}}
\newcommand\cA{{\cal A}}
\newcommand\cS{{\cal S}}
\newcommand\cT{{\cal T}}
\newcommand\cV{{\cal V}}
\newcommand\cL{{\cal L}}
\newcommand\cM{{\cal M}}
\newcommand\cN{{\cal N}}
\newcommand\cG{{\cal G}}
\newcommand\cK{{\cal K}}
\newcommand\cH{{\cal H}}
\newcommand\cI{{\cal I}}
\newcommand\cJ{{\cal J}}
\newcommand\cl{{\iota}}
\newcommand\cP{{\cal P}}
\newcommand\cQ{{\cal Q}}
\newcommand\cg{{\it g}}
\newcommand\cR{{\cal R}}
\newcommand\cB{{\cal B}}
\newcommand\cO{{\cal O}}
\newcommand\tcO{{\tilde {{\cal O}}}}
\newcommand\bz{\bar{z}}
\newcommand\bb{\bar{b}}
\newcommand\ba{\bar{a}}
\newcommand\bg{\bar{g}}
\newcommand\bc{\bar{c}}
\newcommand\bw{\bar{w}}
\newcommand\bX{\bar{X}}
\newcommand\bK{\bar{K}}
\newcommand\bA{\bar{A}}
\newcommand\bZ{\bar{Z}}
\newcommand\bxi{\bar{\xi}}
\newcommand\bphi{\bar{\phi}}
\newcommand\bpsi{\bar{\psi}}
\newcommand\bprt{\bar{\prt}}
\newcommand\bet{\bar{\eta}}
\newcommand\btau{\bar{\tau}}
\newcommand\hF{\hat{F}}
\newcommand\hA{\hat{A}}
\newcommand\hT{\hat{T}}
\newcommand\htau{\hat{\tau}}
\newcommand\hD{\hat{D}}
\newcommand\hf{\hat{f}}
\newcommand\hK{\hat{K}}
\newcommand\hg{\hat{g}}
\newcommand\hp{\hat{\Phi}}
\newcommand\hi{\hat{i}}
\newcommand\ha{\hat{a}}
\newcommand\hb{\hat{b}}
\newcommand\hQ{\hat{Q}}
\newcommand\hP{\hat{\Phi}}
\newcommand\hS{\hat{S}}
\newcommand\hX{\hat{X}}
\newcommand\tL{\tilde{\cal L}}
\newcommand\hL{\hat{\cal L}}
\newcommand\tG{{\tilde G}}
\newcommand\tg{{\tilde g}}
\newcommand\tphi{{\widetilde \Phi}}
\newcommand\tPhi{{\widetilde \Phi}}
\newcommand\te{{\tilde e}}
\newcommand\tk{{\tilde k}}
\newcommand\tf{{\tilde f}}
\newcommand\ta{{\tilde a}}
\newcommand\tb{{\tilde b}}
\newcommand\tc{{\tilde c}}
\newcommand\td{{\tilde d}}
\newcommand\tm{{\tilde m}}
\newcommand\tmu{{\tilde \mu}}
\newcommand\tnu{{\tilde \nu}}
\newcommand\talpha{{\tilde \alpha}}
\newcommand\tbeta{{\tilde \beta}}
\newcommand\trho{{\tilde \rho}}
 \newcommand\tR{{\tilde R}}
\newcommand\teta{{\tilde \eta}}
\newcommand\tF{{\widetilde F}}
\newcommand\tK{{\widetilde K}}
\newcommand\tE{{\widetilde E}}
\newcommand\tpsi{{\tilde \psi}}
\newcommand\tX{{\widetilde X}}
\newcommand\tD{{\widetilde D}}
\newcommand\tO{{\widetilde O}}
\newcommand\tS{{\tilde S}}
\newcommand\tB{{\tilde B}}
\newcommand\tA{{\widetilde A}}
\newcommand\tT{{\widetilde T}}
\newcommand\tC{{\widetilde C}}
\newcommand\tV{{\widetilde V}}
\newcommand\thF{{\widetilde {\hat {F}}}}
\newcommand\Tr{{\rm Tr}}
\newcommand\tr{{\rm tr}}
\newcommand\STr{{\rm STr}}
\newcommand\hR{\hat{R}}
\newcommand\M[2]{M^{#1}{}_{#2}}

\newcommand\bS{\textbf{ S}}
\newcommand\bI{\textbf{ I}}
\newcommand\bJ{\textbf{ J}}

\begin{titlepage}
\begin{center}

\vskip 2 cm
{\LARGE \bf  T-duality of D-brane versus     O-plane  actions
 }\\
\vskip 1.25 cm
  Mohammad R. Hosseini\footnote{Mrhosseiniyazdi@mail.um.ac} and  Mohammad R. Garousi\footnote{garousi@um.ac.ir}

\vskip 1 cm
{{\it Department of Physics, Faculty of Science, Ferdowsi University of Mashhad\\}{\it P.O. Box 1436, Mashhad, Iran}\\}
\vskip .1 cm
 \end{center}

\begin{abstract}

It is known that, in the static gauge,  the world-volume and the transverse Kaluza-Klein (KK)  reductions of the  O-plane effective actions on a circle satisfy  the T-duality constraint for arbitrary base space background. In this paper we show that due to the presence of the second fundamental form in the  D-brane couplings at order $\alpha'$ and higher, the T-duality is  satisfied only for a subclass of the couplings    for arbitrary  base space background. They are $m=0$ couplings where $m$ is the number of $\tB$-field (without derivative on it).     For  $m>0$ couplings, the base space metric must be block-diagonal and the momentum $U(1)$ vector field in the transverse reduction  must be zero. However, the derivatives of the metric and the vector field  are arbitrary.

Using the assumption that the effective actions at the critical dimension are background independent, we then show that the T-duality constraint  for the couplings at order $\alpha'$ and for $m=0$, fixes completely both bulk and  boundary actions.  These couplings  indicate that the propagators of the massless open string  fields  receive $\alpha'$-correction.
 We have also  imposed  the T-duality constraint on $m=1,2,3,4$ couplings. Because of the  above restriction on the base  space background in these cases, however, the T-duality can only partially  fix the couplings for $m>0$. This study shows that the  Dirac-Born-Infeld (DBI) factor appears in both bulk and boundary actions at order $\alpha'$.

\end{abstract}

\end{titlepage}

\section{Introduction}

T-duality is one of the most exciting discoveries in the perturbative  string theory which has been observed  first  in the spectrum  of string  when one compactifies theory  on a circle \cite{Giveon:1994fu,Alvarez:1994dn}. It has been proved in \cite{Sen:1991zi,Hohm:2014sxa} that the KK reduction of the classical effective action of the bosonic and the heterotic string theories on tours $T^d$ are  invariant under $O(d,d)$ transformations  at all orders of $\alpha'$.  Using the assumption that the effective action of string theory at the critical dimension is background independent,  one may consider a particular background which includes a circle. Then one can use the  diffeomorphism symmetry   and the  T-duality symmetry $O(1,1)$  to construct the classical bulk and boundary effective actions of string theory, including their higher derivative corrections \cite{Garousi:2019wgz,Garousi:2019mca,Garousi:2019xlf,Razaghian:2018svg,Garousi:2020gio,Garousi:2020lof,Garousi:2021yyd,Garousi:2021cfc,Garousi:2022ovo}. The T-duality transformations in the base space  are the Buscher rules \cite{Buscher:1987sk,Buscher:1987qj} pluse their  $\alpha'$-corrections \cite{Kaloper:1997ux}, whereas, the  diffeomorphsim and the gauge transformations  in the base space are the standard transformations.   In  the  Double Field Theory formalism in which the T-duality is  imposed before the KK reduction,   however,  the T-duality transformations are the standard $O(D,D)$ transformations whereas the  gauge transformations receive $\alpha'$-corrections \cite{Siegel:1993xq,Hull:2009mi,Marques:2015vua}. 

It has been  also  shown in \cite{Robbins:2014ara,Garousi:2014oya,Akou:2020mxx,Mashhadi:2020mzf}  that the T-duality constraint and the gauge symmetry  can  be used to construct the classical bulk and boundary effective actions of the O-planes of  the superstring theories at order $\alpha'^2$.  These non-perturbative objects  have no dynamics, \ie they have no second fundamental form, however, they  couple  with the closed string states. In this study, one does not consider $\alpha'$-corrections to the Buscher rules which is consistent with the fact that there are no corrections to the T-duality transformations in the type II superstring theory at order $\alpha'^2$.   
  In studying  the  O-plane effective actions, one has to use the static gauge to fix the diffeomorphism symmetry \cite{Robbins:2014ara,Garousi:2014oya}. 

On the other hand, D-branes have open string excitations which describe the D-brane dynamics, and  couple with the closed string states.  It has been shown in \cite{Karimi:2018vaf,Garousi:2022rcv}  that the T-duality and the gauge transformations of the massless open string fields can be used to construct the world-volume couplings of the open string gauge field and the second fundamental form. One can impose the T-duality constraint on the open string fields, either in covariant form \cite{Garousi:2022rcv} or in the static gauge \cite{Karimi:2018vaf}. In this case also  the T-duality transformations   receive $\alpha'$-corrections.  By the invariance under the T-duality, we mean the  T-duality of the  world-volume reduction of D$_p$-brane is the same as the transverse  reduction of D$_{p-1}$-brane, up to some total derivative terms in the base space. The total derivative terms  in turn may fix the couplings on the boundary if the spacetime has boundary and the D-brane is extended in the spacetime such that it ends on the boundary. 

One may use the above T-duality constraint to fix the higher-derivative couplings of the massless closed string fields to the D-brane as well \cite{Garousi:2013gea}.  As in the case of O-plane, one should use the static gauge to fix the diffeomorphism symmetry.   Unlike  the O-plane case, however,   the D-brane couplings include the second fundamental form. In the covariant form, one finds that the world-volume component of the second fundamental form is zero. In the static gauge, however,  this constrains the base space background such that the metric must be block-diagonal and  the momentum  $U(1)$ vector field in the transverse reduction of the  D-brane must be zero. The derivatives of the base space field, however, remain arbitrary.  In the O-plane case, there is no second fundamental form and the $U(1)$ vector field in the transverse reduction  is projected out by the orientifold projections. Under this projection, the base space metric  also becomes block-diagonal. The restriction on the base space background for the D-brane case,  causes  that the T-duality constraint fixes only some of the   independent couplings at order $\alpha'$ and higher.   Those couplings that are fixed by the T-duality constraint should  be  consistent  with the S-matrix elements. This would be  a result of  the assumption that the coefficients of the  independent gauge invariant  couplings  are background independent, \ie if they are fixed in a particular background, they would be  valid for any other arbitrary background as well \cite{Garousi:2022ovo}.

 The massless closed string  $B$-field appears in the world-volume effective action of D-brane  either through its field strength, $H$,  or through  the combination of its pull-back with the massless open string gauge field strength,  $ \tB+F$. Both are  invariant under the $B$-field gauge transformations.   The later appears in the DBI action \cite{Leigh:1989jq,Bachas:1995kx}. This action in the static gauge and only for the  massless closed string fields is
\beqa
S^0_p&=&-T_p\int d^{p+1}\sigma e^{-\Phi}\sqrt{-\det(G_{ab}+  B_{ab})} \labell{a1}
\eeqa
 where $ {T_p} $ is tension of $ {D_p} $-brane, $\Phi$ is the dilaton , $ {{{ G}_{ab}}} $ is the world-volume component of the spacetime metric and $B_{ab}$  is the world-volume component of the $B$-field\footnote{Our index convention is that the Greek letters  $(\mu,\nu,\cdots)$ are  the indices of the space-time coordinates, the Latin letters $(a,d,c,\cdots)$ are the world-volume indices and the Latin letters $(i,j,k,\cdots)$ are the transverse indices. The  tilde indices are the corresponding indices in the base space. The $y$-coordinate is the coordinate of the circle.}. The world-volume KK reduction of D$_p$-brane and the transverse reduction of D$_{p-1}$-brane are, respectively,
\beqa
S^{0w}_p&=&-2\pi T_p\int d^{p}\sigma e^{-\bphi+\vp/4}\sqrt{-\det(\bg_{\ta\tb}+  \bb_{\ta\tb}+g_{[\ta}b_{\tb]}+e^{-\vp}b_{\ta}b_{\tb})}\nn\\
S^{0t}_{p-1}&=&- T_{p-1}\int d^{p}\sigma e^{-\bphi-\vp/4}\sqrt{-\det(\bg_{\ta\tb}+  \bb_{\ta\tb}+b_{[\ta}g_{\tb]}+e^{\vp}g_{\ta}g_{\tb})} \labell{a11}
\eeqa
 where the momentum vector $g$ is resulting from the KK reduction of the spacetime metric and the winding vector $b$ is resulting from the reduction of $B$-field (see equation \reef{GB}). While the above reductions  are not invariant under the $U(1)\times U(1)$ gauge transformations, the transformation of  the world-volume reduction under the Buscher rules,
 \beqa
 \vp\rightarrow -\vp&,& b\leftrightarrow g\labell{buscher}
 \eeqa
 cancels the transverse reduction, \ie they satisfy the T-duality constraint for the most general base space background. 
  Note  that in this case, the T-duality satisfies
  with no residual  total derivative terms which indicates that, as expected,  there is no boundary action at the leading order of $\alpha'$.

The D-brane effective action  at each order of $\alpha'$  has a tower of infinite number of  $ \tB$-field (without derivative on it). Only  at zero-derivative order, this infinite tower of $\tB$-field can appear in the  compact form of the DBI action \reef{a1}. To study the T-duality constraint at the higher orders of $\alpha'$,  one should truncate the couplings at a specific  number of this field, $m$.  As we will see, the  restriction on the base space background discussed earlier,  is only for $m>0$.  In this paper we consider the world-volume couplings of D-brane at order $\alpha'$ in the bosonic string theory in which the $\alpha'$-corrections to the Buscher rules are known \cite{Kaloper:1997ux}.  We will see that at $m=0$ level,  the T-duality satisfies for the arbitrary base space background. In this case, the T-duality constraint fixes completely all independent couplings in the bulk and in the boundary.   At $m>0$ levels, however, because of the restriction on the base space background,  the T-duality constraint can partially fix the independent couplings. These partial results, however,  indicate that both in the bulk and in the boundary actions, there must be  the DBI factor for the couplings at order $\alpha'$. The couplings at $m=0$ level, are fully consistent with the corresponding S-matrix elements. They also   indicate that the standard propagators of the massless open string gauge field and the transverse scalar fields  are corrected at order $\alpha'$.

The outline of the paper is as follows: In section 2, we find all  independent covariant and parity invariant couplings at order $\alpha'$ for $m=0,1,2,3,4$.  For $m=4$ case, we consider only the couplings which include Riemann curvature. We find there are 76 independent couplings.  The coefficients of these couplings are independent of the backgrounds in which  the D$_p$-brane are placed.  In order to  fix the coefficients of  these 76 background independent  couplings and find their corresponding boundary couplings, in section 3, we consider a background which has a boundary. The bulk and  the boundary includes a circle  on which we will impose the T-duality.  In this section we show   the covariant result that the world-volume component of the second fundamental for is zero, is reproduced in the static gauge by requiring the base space background  to satisfy a  particular   condition for $m>0$.  In subsection 3.1, we find that  at $m=0$ level, the T-duality constraint  fixes completely the coefficients of the couplings and their  corresponding boundary coupling.  In subsection 3.2, we show that, for the couplings at $m=1,2,3,4$  levels, the T-duality constraint can not completely fix the parameters of the independent couplings in the bulk and in the boundary. However, the couplings that are found by the T-duality,  indicate that there must be the DBI factor in the couplings at order $\alpha'$ in both bulk and  boundary.    In section 4, we briefly discuss our results. In the Appendix, we show that the  couplings at the $m=0$ level are fully consistent with the corresponding disk-level S-matrix elements. We use the Mathematica package xAct \cite{Nutma:2013zea} for performing the calculations in this paper.

\section{Independent couplings}

We apply the method used in \cite{Garousi:2019cdn} to find the independent couplings.   The independent couplings   are  all covariant and gauge invariant couplings modulo  the field redefinitions,  the Bianchi  identities and the total derivative terms. In this section we are going to find all  independent couplings at order $\alpha'$ which involve the   covariant fields $R$, $H$, $\nabla H$, $\nabla \Phi$, $\nabla\nabla \Phi$,$\tilde{\nabla}\tB$,  $\tilde{\nabla}\tilde{\nabla}\tB$, $\Omega$,  $\tilde{\nabla}\Omega$ and $\tB$. We assume the spacetime curvatures  and all covariant derivatives, except the covariant derivatives of $\tB$, to be constructed with the spacetime metric $G_{\mu\nu}$. The  covariant derivatives of $\tB$ and the world-volume fields are constructed with the pull-back metric $\tG_{ab}$.  The spacetime and the world-volumes  indices should be contracted with the spacetime metric and the with pull-back  metric, respectively.

Each covariant  field may have  world-volume and/or spacetime indices, \eg the field $\tB$ has only world-volume indices
\beqa
\tB_{ab}&\equiv &\prt_a X^\mu\prt_bX^\nu B_{\mu\nu}
\eeqa
Its world-volume derivatives, \eg $\tilde{\nabla}\tB$,  also have only world-volume indices. The $B$-field strength, $H=dB$, on the other hand,  has spacetime and world-volumes indices, \eg $H_{\mu\nu\alpha}$ has only spacetime indices, and $H_{a\mu\nu}$ which is given as
\beqa
H_{a\mu\nu}&\equiv&\prt_aX^\alpha H_{\alpha\mu\nu}
\eeqa
 has both world-volume and spacetime indices.

The second fundamental form has two world-volume and one spacetime indices. It is defined  as the covariant derivative of the world-volume tangent vectors  $\prt_aX^\mu$
\beqa
\Omega^\mu{}_{ab} &\equiv& \tilde{\nabla}_a\prt_b X^\mu\,=\,\prt_a\prt_b X^\mu-\tilde{\Gamma}_{ab}{}^c\prt_c X^\mu+\prt_a X^\nu\prt_b X^\rho \Gamma_{\nu\rho}{}^\mu
\eeqa
where  ${\Gamma}_{\mu\nu}{}^\alpha$ is the  Levi-Civita connection made of the spacetime  metric $G_{\mu\nu}$ and  $\tilde{\Gamma}_{ab}{}^c$ is the  Levi-Civita connection made of the pull-back metric $\tG_{ab}$ which is given as
\beqa
\tG_{ab}&\equiv&\prt_a X^{\mu}\prt_b X^\nu G_{\mu\nu}
\eeqa
Writing  the  Levi-Civita connection  $\tilde{\Gamma}_{ab}{}^c$ in terms of the pull-back metric and using the above relation for  the pull-back metric, one can write the second fundamental form as
\beqa
\Omega^\mu{}_{ab} &=&(G^{\mu\nu}-\tg^{\mu\nu})G_{\nu\alpha}(\prt_a\prt_b X^\alpha+\prt_a X^\nu\prt_b X^\rho \Gamma_{\nu\rho}{}^\alpha)\labell{seconf}
\eeqa
 where  $\tg^{\mu\nu}$ is the first fundamental form which is given as
\beqa
\tg^{\mu\nu}&\equiv &\prt_a X^\mu\prt_b X^\nu \tG^{ab}\labell{firstf}
\eeqa
It is a projection tensor, \ie $\tg^{\mu\nu}G_{\nu\alpha}\tg^{\alpha\beta}=\tg^{\mu\beta}$. It  projects the spacetime tensors to the world-volume directions of the D-brane.  The projection of the second fundamental form along the brane is zero, \ie $\tg^{\mu\nu}G_{\nu\alpha}{}\Omega^\alpha{}_{ab}=0$. Hence, the spacetime index of the second fundamental form can contract only with the spacetime metric, \ie it can not be contracted with the tangent vectors $\prt_aX^\mu$ to convert it to the world-volume index because
\beqa
\Omega^c{}_{ab}&\equiv&\tG^{cd}\prt_dX^\mu\Omega_{ab}{}^{\nu}G_{\mu\nu}=0\labell{O1}
\eeqa
 The   world-volume indices of $\Omega^\mu{}_{ab}$ contract with the inverse of the pull-back metric.

Since  the massless closed string fields are the spacetime fields, one is not allowed to use field redefinitions for these fields in the world-volume effective action of D-brane.  However, as it has been argued in \cite{Robbins:2014ara},  these fields should satisfy the spacetime equations of motion because the D-branes are considered to be probe which do not change the spacetime fields. To impose the equation of motion, we do not consider the Ricci tensor, Ricci scalar, $\nabla_\mu H^{\mu\nu\alpha}$, and $\nabla_\mu\nabla^\mu\Phi$ in the independent couplings. Hence the action should involve all contractions of the following fields:
\beqa
S'_p&=&-\frac{\alpha' T_p}{2}\int d^{p+1}\sigma \, \sqrt { - \det \tilde G_{ab}}\,\cL'(R,H,\nabla\Phi,\Omega,\tilde{\nabla}\tB,\tilde{\nabla}\tilde{\nabla}\tB,\tB )\labell{s1}
\eeqa
where $R$ stands for the Riemann curvature with only  world-volume indices, \ie
\beqa
R_{abcd}&=&\prt_aX^\mu\prt_bX^\nu\prt_cX^\alpha\prt_dX^\beta R_{\mu\nu\alpha\beta}
\eeqa
The world-volume indices in the above Lagrangian should be contracted with $\tG$ and the spacetime indices  should be contracted with the spacetime metric  $G$.  We call the coefficients of the couplings in the above Lagrangian $a_1',a_2',\cdots$.

One has to impose the Bianchi identities in \reef{s1} to find the independent couplings. The Riemann curvature satisfies the following identity:
\beqa
R_{\mu[\nu\alpha\beta]}&=&0 \labell{iden1}
\eeqa
Since there  are no couplings at order $\alpha'$ which involves the derivative of $H$, one does not need to impose the Bianchi identity corresponding to $H$, \ie $dH=0$. However, the couplings do involve the world-volume covariant derivative of the pull-back of the $B$-field. Hence, one should impose its corresponding Bianch identity. One can easily observe the following relation:
\beqa
\tilde{\nabla}_{[a}\tB_{bc]}&=&\prt_{[a}\tB_{bc]}=\prt_a X^\mu\prt_b X^\nu\prt_cX^{\rho}H_{\mu\nu\rho}\,=\, H_{abc}
\eeqa
We consider the scheme in which all couplings involving $H$ in \reef{s1} are  independent, hence, to find the independent couplings involving the derivatives of $\tB$, one has to impose the following Bianchi identity for the   couplings involving $\tilde{\nabla}\tB$:
\beqa
\tilde{\nabla}_{[a}\tB_{bc]}&=&0\labell{iden2}
\eeqa
which is reminiscent to the Bianchi identity of the open string gauge field, \ie $\tilde{\nabla}_{[a}F_{bc]}=0$.

The couplings involving $\tilde{\nabla} \Phi$, $\tilde{\nabla}\tB$, $\tilde{\nabla}\tilde{\nabla}\tB$ and $\tB$ may be related to each others by total derivative terms. The total derivative terms in general should have the following structure:
\beqa
\int d^{p+1}\sigma \sqrt{-\det{\tG_{ab}}}\,\tilde{\nabla}_a(e^{-\Phi}J^a)
\eeqa
where the vector $J^a$ is all contractions of fields  $\tilde{\nabla}\Phi$, $\tilde{\nabla}\tB$  and $\tB$ which involve one derivative. This total derivative relates, among other things, the couplings involving $\tilde{\nabla}\Phi$ to the other couplings involving $\tilde{\nabla}\tB$. We consider  the scheme in which all couplings involving the dilaton in \reef{s1} are independent, so we have to consider the total derivative terms in which the dilaton appears only as the overall factor $e^{-\Phi}$.  Hence we consider the following total derivative terms:
\beqa
\cJ&=&-\frac{\alpha' T_p}{2}\int d^{p+1}\sigma\, e^{-\Phi}\sqrt{-\det{\tG_{ab}}}\,\tilde{\nabla}_a(J^a)
\eeqa
where now the vector $J^a$ is all contractions of fields   $\tilde{\nabla}\tB$ and $\tB$ that involve one derivative. We call their coefficients $j_1, j_2, \cdots$.

If one adds $\cJ$ to the action \reef{s1}, it changes only the coefficients of the gauge invariant couplings $a_1',a_2',\cdots$, \ie
\beqa
S'+\cJ=S\labell{SS}
\eeqa
where $S$ is the same action as \reef{s1} in which the coefficients of the gauge invariant couplings are changed to $a_1,a_2,\cdots$. One can write the above equation as
\beqa
\Delta S+\cJ&=&0\labell{SJK}
\eeqa
where $\Delta S$ is the same as \reef{s1} in which the  coefficients of the gauge invariant couplings are $\delta a_1,\delta a_2,\cdots$ where $\delta a_i=a_i'-a_i$.  To impose the Bianchi identity \reef{iden1}, we write the Riemann curvature in terms of metric and go to the local frame in which the first derivative of metric is zero. To impose the identity \reef{iden2} in the above equation, we rewrite $\tB_{ab}$  in the derivatives of this field  as $\tB_{ab}=\tilde{\nabla}_a A_b-\tilde{\nabla}_b A_a$ where $A_a$ is a vector. Then the above equation involve only independent but non-covariant and non-gauge-invariant couplings. Their coefficients which involve $\delta a_1,\delta a_2,\cdots$ and $j_1,j_2,\cdots$, must be zero.  If one solves the resulting algebraic  equations, one would find some relations between only $\delta a_1,\delta a_2,\cdots$. The number of these relations represents the number of  couplings which are invariant under  the total derivative terms and the Bianchi identities.

Since there can be any number of field  $\tB_{ab}$ in the couplings at any order of $\alpha'$, there  are infinite number of independent couplings at each order of $\alpha'$.  Hence, we have to classify the independent couplings at order $\alpha'$  in sub-structures  in terms of the number of covariant fields $R,\nabla\Phi, \nabla\nabla\Phi, H,\tB,\tilde{\nabla}\tB$,  $\tilde{\nabla}\tilde{\nabla}\tB$, \ie
\beqa
S^1_p&=&-\frac{\alpha' T_p}{2}\int d^{p+1}\sigma \, e^{-\Phi}\sqrt { - \det \tilde G_{ab}}\,\sum_{n=1}^{\infty}L_n\labell{s20}
\eeqa
where $L_n$ is the Lagrangian of independent terms at order $\alpha'$ which have $n$ number of fields $R$, $\nabla\Phi$, $\nabla\nabla\Phi$, $H$, $\tB$, $\tilde{\nabla}\tB$,  $\tilde{\nabla}\tilde{\nabla}\tB$ $\tB$. The number of independent terms in each $L_n$ is fixed, however, there are different schemes that one can choose the independent couplings in each $L_n$. As the number $n$  increases, the number of independent terms in the corresponding $L_n$  also increases. Hence, for performing the calculations we have to truncate the independent couplings at  a specific number $n^*$. In this paper we choose $n^*=6$, \ie we are interested  only in the independent couplings at orders $n=1,2,3,4,5$.

After finding the independent couplings, one may arrange them in terms of the number of
$\tB$-field without derivative on it, \ie
\beqa
S^1_p&=&-\frac{\alpha' T_p}{2}\int d^{p+1}\sigma \, e^{-\Phi}\sqrt { - \det \tilde G_{ab}}\,\sum_{m=0}^{\infty}\cL_m\labell{s2}
\eeqa
where $\cL_m$ is the Lagrangian of the independent terms found in \reef{s20} which have $m$ field $\tB$.  The couplings in \reef{s20} for $n=1,2,3,4,5$ produces $\cL_m$ with  $m=0,1,2,3,4$. We have found the following 11 independent terms for $m=0$ case in a particular scheme:
\beqa
\cL_0&=&b_{11}H_{abc} H^{abc} + b_{4}H_{ab\mu} H^{ab\mu} +b_{2}H_{a\mu \nu } H^{a\mu \nu } +
b_1 H_{\mu \nu \rho } H^{\mu \nu \rho } + b_{5}R ^{ab}{}_{ab}\nn\\
 && +  a_{1} \Omega_{\mu }{}^{b}{}_{b} \Omega^{\mu a}{}_{a} + a _ 2 \Omega_{\mu ab} \Omega^{\mu ab} +  a_{8}\tilde{\nabla}_{a}\tB_{bc} \tilde{\nabla}^{a}\tB^{bc} \nn\\&&+
b_{47}\nabla_{a}\Phi \nabla^{a}\Phi + b _{42}\Omega_{\mu }{}^{a}{}_{a} \nabla^{\mu }\Phi +b_{45}\nabla_{\mu }\Phi \nabla^{\mu }\Phi \labell{act1}
\eeqa
where the world-volume indices are contracted with the inverse of the pull-back metric $\tG_{ab}$ and the spacetime indices are contracted with the spacetime metric $G_{\mu\nu}$ and its inverse $G^{\mu\nu}$. Apart from the couplings with coefficient $a_8$, all other couplings are the independent couplings that have been considered in \cite{Garousi:2013gea}. In that paper, the second fundamental form is defined as
\beqa
\Omega'^\mu{}_{ab}&=& \prt_a\prt_b X^\alpha+\prt_a X^\nu\prt_b X^\rho \Gamma_{\nu\rho}{}^\alpha
\eeqa
with the assumption that  its spacetime index is contracted with the transverse projection operator $\bot_{\mu\nu}=G_{\mu\nu}-\tg_{\mu\nu}$. It is the same as the second fundamental form \reef{seconf} with the assumption that its spacetime index is contracted with the spacetime metric $G_{\mu\nu}$.

We have also found the following 6, 35, 17 and 7 independent terms, in a particular scheme, for $m=1,2,3,4$, respectively:
\beqa
\cL_1&=& b_{27}  \tB^{ab} H_{bc\mu} \Omega^{\mu }{}_{a}{}^{c} +  b_{28}   \tB^{ab} H_{ab\mu } \Omega^{\mu c}{}_{c} + b_{36 }  \tB^{bc} H_{abc} \nabla^{a}
\Phi \nn\\&&-   b_{ 50 } \tB^{bc} \nabla^{a}\Phi\tilde{\nabla}_{b}\tB_{ac} +  b_{51}   \tB_{a}{}^{b} \nabla^{a}\Phi \tilde{\nabla}^{c}\tB_{bc} + b_{35} \tB^{ab} H_{ab\mu } \nabla^{\mu }\Phi\nn\\
\cL_2&=& b_{ 19}
\tB^{ab} \tB^{cd} H_{ac}{}^{e}
H_{bde} +
b_{ 7}
\tB^{ab} \tB^{cd} H_{ac}{}^{\mu } H_{bd\mu } +
 b_{ 20}
\tB^{ab} \tB^{cd} H_{ab}{}^{e} H_{cde} +
b_{ 21}
\tB_{a}{}^{c} \tB^{ab}
H_{b}{}^{de} H_{cde} \nn\\&&+
b_{ 8}
\tB^{ab} \tB^{cd} H_{ab}{}^{\mu } H_{cd\mu } +
 b_{ 12}
\tB_{a}{}^{c} \tB^{ab} H_{b}{}^{d\mu } H_{cd\mu }
+
b_{ 6}
\tB_{a}{}^{c} \tB^{ab}
H_{b}{}^{\mu \nu } H_{c\mu \nu } +
 b_{ 22}
\tB_{ab} \tB^{ab} H_{cde} H^{cde}\nn\\&& +
b_{ 14}
\tB_{ab} \tB^{ab} H_{cd\mu }
H^{cd\mu } +
b_{ 10}
\tB_{ab} \tB^{ab} H_{c\mu \nu } H^{c\mu \nu } +
 b_{ 3}
\tB_{ab} \tB^{ab} H_{\mu \nu \rho } H^{\mu \nu
\rho } +
b_{ 9}
\tB^{ab} \tB^{cd} R _{acbd} \nn\\&&+
 b_{ 13}
\tB_{a}{}^{c} \tB^{ab} R _{b}{}^{d}{}_{cd}
+
b_{ 15}
\tB_{ab} \tB^{ab} R
^{cd}{}_{cd} +
a_{ 5}
\tB^{ab} \tB^{cd} \Omega_{\mu bd} \Omega^{\mu }{}_{ac} +
a_{ 3}
\tB_{a}{}^{c} \tB^{ab}
\Omega_{\mu }{}^{d}{}_{d} \Omega^{\mu }{}_{bc} \nn\\&&+
a_{ 6}
\tB_{a}{}^{c} \tB^{ab}
\Omega_{\mu cd} \Omega^{\mu }{}_{b}{}^{d} +
 a_{ 4}
\tB_{ab} \tB^{ab} \Omega_{\mu }{}^{d}{}_{d}
\Omega^{\mu c}{}_{c} +
a_{ 7}
\tB_{ab} \tB^{ab} \Omega_{\mu cd} \Omega^{\mu cd} +
b_{ 49}
\tB_{bc} \tB^{bc}
\nabla_{a}\Phi \nabla^{a}\Phi \nn\\&&+
 b_{ 48}
\tB_{a}{}^{c} \tB_{bc} \nabla^{a}\Phi
\nabla^{b}\Phi +
a_{ 12}
\tB_{ab} \tB^{ab} \tilde{\nabla}_{c}\tB_{de} \tilde{\nabla}^{c}\tB^{de} -
b_{ 54}
\tB_{a}{}^{c} \tB^{ab} H_{cde}
\tilde{\nabla}^{d}\tB_{b}{}^{e} \nn\\&&+
a_{ 11}
\tB_{a}{}^{c} \tB^{ab} \tilde{\nabla}_{d}\tB_{ce} \tilde{\nabla}^{d}\tB_{b}{}^{e} +
a_{ 9}
\tB_{a}{}^{c} \tB^{ab}
\tilde{\nabla}^{d}\tB_{b}{}^{e} \tilde{\nabla}_{e}\tB_{cd} +
 b_{ 52}
\tB^{ab} \tB^{cd} H_{cde} \tilde{\nabla}^{e}\tB_{ab} \nn\\&&+
a_{ 13}
\tB^{ab} \tB^{cd}
\tilde{\nabla}_{e}\tB_{cd} \tilde{\nabla}^{e}\tB_{ab} +
 b_{ 53}
\tB^{ab} \tB^{cd} H_{bde} \tilde{\nabla}^{e}\tB_{ac} +
a_{ 10}
\tB^{ab} \tB^{cd}
\tilde{\nabla}_{e}\tB_{bd} \tilde{\nabla}^{e}\tB_{ac} \nn\\&&+
 b_{ 55}
\tB^{ab} \tB^{cd} H_{bcd} \tilde{\nabla}^{e}\tB_{ae} +
a_{ 14}
\tB_{a}{}^{c} \tB^{ab}
\tilde{\nabla}_{b}\tB_{c}{}^{d} \tilde{\nabla}^{e}\tB_{de} +
 a_{ 15}
\tB_{ab} \tB^{ab} \tilde{\nabla}^{c}\tB_{c}{}^{d}
\tilde{\nabla}^{e}\tB_{de} \nn\\&&+
b_{ 43}
\tB_{a}{}^{c} \tB^{ab} \Omega_{\mu bc} \nabla^{\mu }\Phi +
b_{ 44}
\tB_{ab} \tB^{ab}
\Omega_{\mu }{}^{c}{}_{c} \nabla^{\mu }\Phi +
b_{ 46}
\tB_{ab} \tB^{ab}
\nabla_{\mu }\Phi \nabla^{\mu }\Phi \nn\\
\cL_3&=& b_{
 29 }
  \tB_{a}{}^{c} \tB^{ab} \tB^{de}
H_{de\mu } \Omega^{\mu }{}_{bc} +
  b_{
30 }
 \tB_{a}{}^{c} \tB^{ab} \tB^{de} H_{ce\mu }
\Omega^{\mu }{}_{bd} +
 b_{  31
}
\tB_{a}{}^{c} \tB^{ab} \tB_{b}{}^{d} H_{de\mu } \Omega^{\mu
}{}_{c}{}^{e} \nn\\&&+
 b_{  32
}
\tB_{ab} \tB^{ab} \tB^{cd} H_{de\mu } \Omega^{\mu }{}_{c}{}^{e} +
b_{
 33 }
  \tB_{a}{}^{c} \tB^{ab}
\tB_{b}{}^{d} H_{cd\mu } \Omega^{\mu e}{}_{e} +
 b_{
 34 }
  \tB_{ab} \tB^{ab} \tB^{cd} H_{cd
\mu } \Omega^{\mu e}{}_{e}\nn\\&& +
  b_{
39 }
 \tB_{b}{}^{d} \tB^{bc} \tB_{c}{}^{e} H_{ade}
\nabla^{a}\Phi +
 b_{  40
}
\tB_{bc} \tB^{bc} \tB^{de} H_{ade} \nabla^{a}\Phi +
  b_{
41 }
 \tB_{a}{}^{b} \tB_{b}{}^{c} \tB^{de} H_{cde}
\nabla^{a}\Phi \nn\\&&-
 b_{  56
}
\tB_{b}{}^{d} \tB^{bc} \tB_{c}{}^{e} \nabla^{a}\Phi \tilde{\nabla}_{d}\tB_{ae} -
b_{
 58 }
  \tB_{bc} \tB^{bc} \tB^{de}
\nabla^{a}\Phi \tilde{\nabla}_{d}\tB_{ae} +
 b_{
57 }
 \tB_{a}{}^{b} \tB_{c}{}^{e} \tB^{cd} \nabla^{a}\Phi
\tilde{\nabla}_{d}\tB_{be}\nn\\&& -
 b_{  59
}
\tB_{a}{}^{b} \tB_{b}{}^{c} \tB^{de} \nabla^{a}\Phi \tilde{\nabla}_{d}\tB_{ce} +
b_{
 60 }
  \tB_{a}{}^{b} \tB_{cd} \tB^{cd}
\nabla^{a}\Phi \tilde{\nabla}^{e}\tB_{be} +
  b_{
61 }
 \tB_{a}{}^{b} \tB_{b}{}^{c} \tB_{c}{}^{d}
\nabla^{a}\Phi \tilde{\nabla}^{e}\tB_{de} \nn\\&&+
  b_{
37 }
 \tB_{a}{}^{c} \tB^{ab} \tB_{b}{}^{d} H_{cd\mu }
\nabla^{\mu }\Phi +
 b_{  38
}
\tB_{ab} \tB^{ab} \tB^{cd} H_{cd\mu } \nabla^{\mu }\Phi \nn\\
\cL_4&=& b_{
 16 }
  \tB_{a}{}^{c} \tB^{ab}
\tB_{d}{}^{f} \tB^{de}R _{becf} +
 b_{
 17 }
  \tB_{a}{}^{c} \tB^{ab}
\tB_{b}{}^{d} \tB^{ef}R _{cdef} +
 b_{
 18 }
  \tB_{ab} \tB^{ab} \tB^{cd}
\tB^{ef}R _{cedf}\nn\\&& +
b_{ 23
}
\tB_{a}{}^{c} \tB^{ab} \tB_{b}{}^{d} \tB_{c}{}^{e}R
_{d}{}^{f}{}_{ef} +
b_{ 24
}
\tB_{ab} \tB^{ab} \tB_{c}{}^{e} \tB^{cd}R _{d}{}^{f}{}_{ef} +
b_{
 25 }
  \tB_{a}{}^{c} \tB^{ab}
\tB_{b}{}^{d} \tB_{cd}R ^{ef}{}_{ef} \nn\\&&+
 b_{
26 }
 \tB_{ab} \tB^{ab} \tB_{cd} \tB^{cd}R
^{ef}{}_{ef}\labell{act2}
\eeqa
where $a_1,\cdots a_{15}$ and $b_1,\cdots b_{61}$ are background independent coefficients. There are also many couplings involving $H^2, H\tilde{\nabla}\tB, (\tilde{\nabla}\tB)^2, (\nabla\Phi)^2$ in $\cL_4$ above. They are six-field couplings that   we did not consider in \reef{s20}.  Note that the couplings in $\cL_2$ with coefficients $b_{22}$,  $b_{14}$,  $b_{10}$,  $b_3$,  $b_{15}$, $a_4$,  $a_7$,  $b_{49}$,  $a_{12}$,  $b_{44}$,  $b_{46}$ are the same as the couplings in $\cL_0$ with one extra factor $\tB^{ab}\tB_{ba}$.  The couplings  in $\cL_3$ with coefficients $b_{32}$, $b_{34}$, $b_{40}$, $b_{58}$, $b_{60}$, $b_{38}$ are the same as the couplings in $\cL_1$ with one extra factor $\tB^{ab}\tB_{ba}$. The coupling in $\cL_4$ with coefficients $b_{25}$,  $b_{26}$ are the same as the Riemann curvature coupling in $\cL_0$ with the extra factors  $\tB^{ab}\tB_{bc}\tB^{cd}\tB_{da} $ and  $(\tB^{ab}\tB_{ba})^2$, respectively. We will see in the next section that most of these coefficients  are fixed by the T-duality constraint. The result is such that they can be   reproduced by  the overall DBI factor for the couplings at order $\alpha'$.

Using  the replacement $\tB\rightarrow \tB+F$, the above independent couplings are covariant and are invariant under the $B$-field and the massless open string  gauge transformations. To study these transformations under the T-duality constraint, one has to use the static gauge in which $X^a=\sigma^a$. In this gauge, the transverse spacetime coordinates $X^i$ are the open string transverse scalar fields that describe the dynamics of the D-branes.

In the static gauge and for zero open string fields, all the above couplings have  closed string contributions, so they  may be fixed by imposing the T-duality constraint on the closed string fields. For zero closed string fields, the couplings with coefficients $a_1,\cdots a_{15}$ have open string contributions, so they may also be fixed by considering the T-duality of the open string fields. They have been studied in \cite{Karimi:2018vaf} when one uses the field redefinitions for the open string fields. In that case, not all  15 couplings are independent. In fact,  the number of independent parameters reduces to 7 parameters \cite{Karimi:2018vaf} which can also be fixed by comparing them with the low energy expansion of  the disk-level S-matrix element of four gauge boson vertex operators.  However, if one does not use the field redefinitions, one would find 15 independent couplings which  in a particular scheme, they are the same as the above couplings with  coefficients $a_1,\cdots a_{15}$ in which $\tB$ is replaced by $F$.  In the static gauge, all   couplings  have open and closed string contributions, so they  may also be fixed by considering  the T-duality  of massless open and closed string fields. In this paper, however,    we are going to see to what extend  the T-duality of the massless closed string fields fixes these parameters.

\section{T-duality constraint}

 There is a theorem that indicates the classical spacetime effective action of the bosonic string theory at any order of $\alpha'$ should be  invariant under the T-duality when one compactifies the theory on a tours \cite{Sen:1991zi}. There is no restriction on the  base space background in this study because the couplings in the base space are covariant. The T-duality transformations should be    the Buscher rules plus some specific $\alpha'$-corrections \cite{Kaloper:1997ux} which depend on the scheme of the independent spacetime couplings \cite{Garousi:2019wgz}.  To study the T-duality of the closed string couplings of the D-brane effective action, however, one should fix the diffeomorphism by choosing the static gauge.   There is no theorem that indicates  the D-branes effective actions at any order of $\alpha'$, in the static gauge, should be invariant under the T-duality for arbitrary base space background.   In this section  we are going to show that at the $m=0$ level, the base space background is arbitrary, whereas at the  $m>0$ levels, the background should be specific such that metric in the base space must be block-diagonal and the  gauge field $g_\ta$ in the transverse reduction must be  zero. There is, however,  no restriction on the derivatives of the base space fields.     If one could fix the couplings in the actions \reef{act1} and \reef{act2} for such a specific background, then the background independence assumption guaranties that they are valid for any other arbitrary  background as well.

We are interested in the massless closed string couplings in  the actions \reef{act1} and \reef{act2}, hence, one should fix the world-volume diffeomorphism  by using the static gauge and for $X^i=0$.  In this gauge, the pull-back of metric and $B$-field are
\beqa
\tG_{ab}\,=\,G_{ab}&;& \tB_{ab}\,=\, B_{ab}
\eeqa
The first fundamental form \reef{firstf} has only world-volume components, \ie
\beqa
\tg^{ij}=\tg^{ai}=\tg^{ia}=0&;&\tg^{ab}=\tG^{ab}
\eeqa
In the static gauge, the second fundamental form satisfies   the  identity \reef{O1} provided that
\beqa
\tG^{ab}G_{b\mu}&=&\delta^a_\mu\labell{inverstG}
\eeqa
 To verify the above relation, we calculate $\Omega^c{}_{ab}$ from equation \reef{seconf}  in the static gauge and for $X^i=0$, \ie
\beqa
\Omega^c{}_{ab}&=& \Gamma_{ab}{}^c-\tG^{cd}G_{d\nu}\Gamma_{ab}{}^\nu
\eeqa
which becomes  zero only when the relation \reef{inverstG} is satisfied\footnote{
The identity \reef{O1} for the constraint \reef{inverstG} in the static gauge   becomes
\beqa
\tG^{cd}\prt_d X^i \Omega^j{}_{ab} G_{ij}&=&0
\eeqa
For the flat spcetime metric,  one finds $\Omega^i{}_{ab}=\prt_a\prt_b X^i$. The above relation becomes
\beqa
\prt_c X^i\prt_a\prt_b X^j\eta_{ij}&=&0
\eeqa
Hence,  the derivative of the pull-back metric becomes zero. Therefore, in the static gauge,  the derivatives for the massless open string fields become only partial derivatives.}.
The transverse component of the second fundamental form in the static gauge and for $X^i=0$ becomes  $\Omega^i{}_{ab}=\Gamma_{ab}{}^i$. Note that in order to satisfy the constraint \reef{inverstG}, one may consider the spacetime metric which is block-diagonal, \ie $G_{ai}=0$. However, such metric would  not be consistent with  the KK reduction of the spacetime metric that we are going to use in this paper,  \ie it would not be consistent with \reef{GB}.

To study the T-duality,  we begin with  a specific background which has a circle and an arbitrary base space. That is, the   manifold has the structure $M^{(26)} = M^{(25)} \times S^{(1)}$.
 The manifold $M^{(26)}$ has coordinates $x^\mu = (x^{\tilde{\mu}}, y)$ where $x^{\tilde{\mu}}$ is the coordinates of the base space manifold $M^{(25)}$, and $y$ is the coordinate of the circle $S^{(1)}$.
 The KK   reduction of  the spacetime metric $ G_{\mu\nu} $,  $ B_{\mu\nu} $ and the dilaton $\Phi$ are \cite{Maharana:1992my}
\beqa
G_{\mu\nu}=
\left(\matrix{
\bg_{\tmu \tnu}+e^{\varphi}g_\tmu g_\tnu   &  e^{\varphi}g_\tmu &\cr
e^{\varphi}g_\tnu  &  e^{\varphi}&}\!\!\!\!\!\right)
; B_{\mu\nu}=\left(\matrix{
\bar{b}_{\tmu \tnu}-\frac{1}{2} g_\tmu b_\tnu+\frac{1}{2}g_\tnu b_\tmu   &   b_\tmu&\cr
-b_\tnu & 0&
}\!\!\!\!\!\right);\Phi=\bphi+\vp/4\labell{GB}
\eeqa
Inverse of the spacetime  metric is
\beqa
G^{\mu\nu}=
\left(\matrix{
\bg^{\tmu \tnu} & -g^\tmu &\cr
-g^\tnu & e^{-\varphi}+g_\talpha g^\talpha &
}\right)\labell{IG}
\eeqa
Using these reductions, it is straightforward to calculate the reduction of the spacetime tensors
 $ R_{\mu\nu\rho\sigma} $,  $ H_{\mu\nu\rho} $,  and $ \nabla_\mu \Phi $ which appear in the couplings \reef{act1} and \reef{act2}. However, these actions have world-volume fields as well. The reduction of the world-volume fields dependents on position of   the D-brane in the spacetime.

When D$_p$-plane is along the $y$-direction, \ie $M^{(26)}=M^{(p+1)}\times M^{(25-p)}$ and $M^{(p+1)}=S^{(1)}\times M^{(p)}$,  the reduction of pull-back metric  $ \tG_{ab} $ and its inverse   in the static gauge are
\beqa
\tG_{ab}=
\left(\matrix{
{\bg}_{\tilde{a} \tilde{b}}+e^{\varphi}{g}_{\tilde{a}}{g}_{\tilde{b}} & e^{\varphi}{g}_{\tilde{a}}&\cr
e^{\varphi}{g}_{\tilde{b}} & e^{\varphi}&
}\right)\qquad;\qquad \tG^{ab}=
\left(\matrix{
\bg^{\tilde{a} \tilde{b}} & -g^{\tilde{a}} &\cr
-g^{\tilde{b}} & e^{-\varphi}+g_{\tilde{c}} g^{\tilde{c} }&
}\right)\labell{wwR}
\eeqa
where the indices $\ta,\tb$ are world-volume indices that do not include the world-volume index $y$, \ie they are belong to $M^{(p)}$. In above equation, the indices are raised by $\bg^{\ta\tb}$, and   $\bg^{\ta\tb}$ is inverse of $\bg_{\ta\tb}$. Using the reductions \reef{GB} and \reef{wwR}, one can easily verify that the constraint  \reef{inverstG} is satisfied for the world-volume reduction when the base space  metric is block-diagonal, \ie.
\beqa
\bg_{\tmu\tnu}=
\left(\matrix{
{\bg}_{\tilde{a} \tilde{b}} &0&\cr
0 & {\bg}_{\tilde{i} \tilde{j}}&
}\right)\labell{bd}
\eeqa
Note that while the consistency for the second fundamental form requires $\bg_{\ta i}$ to be zero,  there is no constraint on the derivatives of this field. The reduction of the overall factor in the action \reef{s2} in this case is $e^{-\Phi}\sqrt{-\tG}=e^{-\bphi+\vp/4}\sqrt{-\bg}$.

When D$_{(p-1)}$-plane is orthogonal to  the $y$-direction, \ie  $M^{(26)}=M^{(p)}\times M^{(26-p)}$ and $M^{(26-p)}=S^{(1)}\times M^{(25-p)}$,  the reduction of the pull-back metric and its inverse  are
\beqa
\tG_{\ta\tb}\,\,=\,\,\bg_{\ta\tb}+e^{\varphi}{g}_{\ta}g_{\tb}&;& \tG^{\ta\tb}\,\,=\,\,\bg^{\ta\tb}-\frac{g^\ta g^\tb}{e^{-\varphi}+g_\tc g^\tc}\labell{tR}
\eeqa
 In this case, however, the  constraint  \reef{inverstG}  is satisfied  for the case that the base space  metric is block-diagonal \reef{bd} and  the vector $g_\ta$ is zero. In fact, for the block-diagonal base space metric, one finds
 \beqa
 \tG^{\ta\tb}G_{\tb\tc}\,=\,\delta^\ta_{\tc}\,\, ,\,\, \tG^{\ta\tb}G_{\tb \tilde{i}}\,=\,\frac{g^\ta g_{\tilde{i}}}{e^{-\vp}+g_\tc g^\tc}\,\, ,\,\, \tG^{\ta\tb}G_{\tb y}\,=\,\frac{g^\ta }{e^{-\vp}+g_\tc g^\tc}
 \eeqa
 which satisfy the constraint  \reef{inverstG} for $g_\ta=0$. Here also we note that the consistency for the second fundamental form in the static gauge requires  no constraint on the derivatives of this field.   For the couplings  at $m=0$ level,  as we will see shortly, in imposing the T-duality constraint one has to consider  one and two base space fields. On the other hand, the  second fundamental form does not appear in the effective action \reef{act1} solely. Hence, in the T-duality at order $m=0$ level, one has to consider the linear terms of the second fundamental form. At the linear order the relation $\Omega^c{}_{ab}=0$ is satisfied with no constraint on the base space background.   Hence, in the transverse reduction of the  $m>0$ couplings, the consistency with the   covariant identity \reef{O1} requires the base space in the static gauge to be restricted to \reef{bd} and
 \beqa
 g_{\ta}\,=\,0&;& \prt g_{\ta}\,\neq 0\labell{ta0}
 \eeqa
Note that there is no constraint on the vector $g_\ta$ in the world-volume reduction.  There is no constraint on the background for $m=0$ case. The transverse  reduction of the overall factor in this case is $e^{-\Phi}\sqrt{-\tG}=e^{-\bphi-\vp/4}\sqrt{-\bg}$. Note that   the second fundamental form does not  appear in the leading order DBI action, hence, in studying the KK reductions of the DBI action in \reef{a11} one does not need to impose the above restriction on the base space background. Note also that for the O-plane case,  the vector $g_{\ta}$ and the metric $\bg_{\ta\tilde{i}}$ are removed by the orientifold projection. Hence, in that case also the base space background is arbitrary.

   When the $D_p$-brane is along the circle, using the reductions \reef{GB} and \reef{wwR},  it is straightforward to calculate  the world-volume reduction of $S_p^1$. We  call it $S_p^{1w}$. When the $D_p$-brane is orthogonal to the circle,  using the reductions \reef{GB} and \reef{tR}, it is straightforward to calculate  the transverse  reduction of $S_p^1$. We  call it $S_p^{1t}$.  These two actions are not identical. However, the transformation   of $S_p^{1w}$ under the  Buscher rules
 which is called $S_{p-1}^{1wT}$, may  be the same as $S_{p-1}^{1t}$, up to some total derivative terms  in the base space, \ie
 \beqa
\Delta \tilde{S}^1_0+\tilde{\cJ}&=&0\labell{tSJK}
\eeqa
where  $\Delta \tilde{S}^1_0 = S_{p-1}^{1wT}-S_{p-1}^{1t} $ and  the total derivative term
is
\beqa
\tilde{\cJ}&=&-\frac{\alpha' T_{p-1}}{2}\int d^{p}\sigma\, \sqrt { - \bg}\,\nabla_{\tilde{a}}(e^{-\bphi-\vp/4}{\cI}^{\tilde{a}})\labell{ttot}
\eeqa
where ${\cI}^{\tilde{a}}$ is a  vector which is made of the base space fields and their  derivatives at order $\alpha'^{1/2}$ with coefficients $I_1,I_2,\cdots$.  It has been observed in \cite{Garousi:2019mca} that the T-duality constraint for the flat base space background produces the same constraints as for the  curved base space background. Hence, for simplicity of the calculations we consider the base space metric to be flat. For this base space matric, the constraint \reef{bd}  is satisfied. The T-duality constraint \reef{tSJK} should be satisfied if the T-duality transformations are only the Buscher rules. The subscript 0 in  $\Delta \tilde{S}^1_0$ refers to the T-duality transformations at order $\alpha'^0$. However, if the T-duality transformations receive $\alpha'$-corrections then there should be some other terms in \reef{tSJK}  resulting from the T-duality of the leading order action.

In fact the T-duality transformations do have $\alpha'$-corrections. At order $\alpha'$, they  have been found in \cite{Kaloper:1997ux}. They are
\beqa
\vp&\rightarrow& -\vp-\alpha' \lambda_0\bigg[2\nabla_\tmu\vp\nabla^\tmu\vp+e^{\vp}V_{\tmu\tnu}V^{\tmu\tnu}+e^{-\vp}W_{\tmu\tnu}W^{\tmu\tnu}\bigg]\nn\\
g_{\tmu}&\rightarrow& b_{\tmu}-\alpha' \lambda_0\bigg[2W_{\tmu\tnu}\nabla^{\tnu}\vp+e^{\vp}\bar{H}_{\tmu\tnu\talpha}V^{\tnu\talpha}\bigg]\nn\\
b_{\tmu}&\rightarrow& g_{\tmu}-\alpha' \lambda_0\bigg[2V_{\tmu\tnu}\nabla^{\tnu}\vp-e^{-\vp}\bar{H}_{\tmu\tnu\talpha}W^{\tnu\talpha}\bigg]\nn\\
\bar{b}_{\tmu\tnu}&\rightarrow& \bar{b}_{\tmu\tnu}-\alpha' \lambda_0\bigg[4V_{\talpha[\tmu}W^\talpha{}_{\tnu]}+2g_{[\tnu}W_{\tmu]\talpha}\nabla^\talpha\vp+2b_{[\tnu}V_{\tmu]\talpha}\nabla^\talpha\vp\nn\\
&&\qquad\qquad\quad+e^{\vp}g_{[\tnu}\bar{H}_{\tmu]\talpha\tbeta}V^{\talpha\tbeta}-e^{-\vp}b_{[\tnu}\bar{H}_{\tmu]\talpha\tbeta}W^{\talpha\tbeta}\bigg]
\labell{a20}
\eeqa
where  $\lambda_0=-1/4$ for the bosonic string theory. In the above transformations, $ \bar{H}_{\tmu\tnu\trho}\equiv 3\partial_{[\tmu}\bar{b}_{\tnu\trho]}-\frac{3}{2}g_{[\tmu}W_{\tnu\trho]}-\frac{3}{2}b_{[\tmu}V_{\tnu\trho]} $,  $ V_{\tmu\tnu}=\partial_{\tmu}g_{\tnu}-\partial_{\tnu}g_{\tmu} $ and $ W_{\tmu\tnu}=\partial_{\tmu}b_{\tnu}-\partial_{\tnu}b_{\tmu} $ are the  field strengths in the base space. In the reductions of the spacetime effective actions and the O-plane effective actions, only these field strengths appear in the base space, \ie the couplings  and the total derivative terms are all gauge invariant. However, the effective actions of  the D-brane involve $\tB_{ab}$, hence, the base space fields $\bar{b}_{\ta\tb}, g_\ta,b_\ta$ with and without  derivative on them appear in the reductions of the couplings. Therefore, the reductions of the couplings  in the base space should be in terms of  the base space fields $\bar{b}_{\ta\tb}, g_\ta,b_\ta$ and their derivatives rather than the field strengths $\bar{H},W,V$. Therefore, ${\cI}^{\tilde{a}}$ in \reef{ttot} should be a  vector which is made of the base space fields $\bar{b}_{\ta\tb}, g_\ta,b_\ta,\vp,\bphi$ and their  derivatives at order $\alpha'^{1/2}$.

Since the T-duality transformations are the Buscher rules \reef{buscher} plus  the above $\alpha'$-corrections,  the DBI reductions \reef{a11} satisfy the T-duality only at order $\alpha'^0$, \ie $\Delta\tS^0_0=0$. At order $\alpha'$ it produces the following residual terms:
\beqa
\Delta\tS^0_1&=&- 2\pi\alpha'T_{p}\int d^{p}\sigma e^{-\bphi-\vp/4}\sqrt{-\det(A^0_{\ta\tb})} \bigg[\frac{1}{4}\Delta\vp+\frac{1}{2}\Tr[(A^0)^{-1}A^1]\bigg]\labell{DS0}
\eeqa
where
\beqa
A^0_{\ta\tb}&=&\bg_{\ta\tb}+  \bb_{\ta\tb}+b_{[\ta}g_{\tb]}+e^{\vp}g_{\ta}g_{\tb}\nn\\
A^1_{\ta\tb}&=&\Delta\bar{b}_{\ta\tb}+b_{[\ta}\Delta b_{\tb]}+g_{[\ta}\Delta g_{\tb]}+2e^{\vp}g_{\{\ta}\Delta b_{\tb\}}-e^{-\vp}g_{\ta}g_{\tb}\Delta\vp
\eeqa
and we have written the T-duality transformations \reef{a20} as
\beqa
\vp\rightarrow -\vp+\alpha'\Delta\vp\,\,, g_{\ta}\rightarrow b_{\ta}+\alpha'\Delta g_{\ta}\,\,,b_{\ta}\rightarrow g_{\ta}+\alpha'\Delta b_{\ta}\,\,,\bar{b}_{\ta\tb}\rightarrow \bar{b}_{\ta\tb}+\alpha'\Delta \bar{b}_{\ta\tb}
\eeqa
The above terms should be included in the T-duality constraint \reef{tSJK}, \ie
\beqa
\Delta\tS^0_1+\Delta\tS^1_0+\tilde{\cJ}&=&0\labell{cons}
\eeqa
The above constraint should fix the parameters in the actions \reef{act1} and \reef{act2}.  Note that there is no parameter in the first term of \reef{cons}. To study the constraint \reef{cons}, one should expand the DBI contribution \reef{DS0} and the reductions of  the actions \reef{act1} and \reef{act2} in terms of the number of the base space fields  $\bar{b}_{\ta\tb}, g_\ta,b_\ta,\prt\vp,\prt\bphi$, and their derivatives.

The  DBI contribution $\Delta\tS^0_1$, $\Delta\tS^1_0$ and the total derivative term \reef{ttot} have the expansions
\beqa
\Delta\tS^0_1&=&\sum_{\tm=2}^{\infty}\Delta\tS^0_1(\tm)\nn\\
\Delta\tS^1_0&=&\sum_{\tm=1}^{\infty}\Delta\tS^1_0(\tm)\nn\\
\tilde{\cJ}&=&\sum_{\tm=1}^{\infty}\tilde{\cJ}(\tm)
\eeqa
where $\tm$ is the number of the  base space fields  $\bar{b}_{\ta\tb}, g_\ta,b_\ta,\prt\vp,\prt\bphi$  and their derivatives.  For $\tm>2$, one should also impose the constraint \reef{ta0} on the base space background. In principle, if one considers the D-brane action \reef{s2} to include  all infinite independent terms at order $\alpha'$,  then the constraint  \reef{cons} would  produce infinite relations between their parameters. In practice, however, one should consider the independent couplings in \reef{s2} up to a specific number $m$. Then in the above expansion one should consider the terms up to $\tm=m+1$ for $m>0$. For $m=0$ case, one should consider the terms up to $\tm=2$.

Since there is no restriction on the base space background for the $m=0$ case, as we will see in the next subsection, the constraint \reef{cons} produces so many relations between the parameters in Lagrangian \reef{act1} which fix the bulk and boundary actions completely. However, to be able to study the T-duality constraint for $m>0$ cases, the base space background should be  restricted as in \reef{ta0}. This causes that the constraint \reef{cons} does not produce so many relations between the parameters in Lagrangian \reef{act2} to fix the bulk and boundary actions completely.

\subsection{$m=0$ case}

The T-duality constraint for $m=0$ case is
\beqa
\sum_{\tm=2}^{2}\Delta\tS^0_1(\tm)+\sum_{\tm=1}^{2}\Delta\tS^1_0(\tm)+\sum_{\tm=1}^{2}\tilde{\cJ}(\tm)&=&0\labell{mm}
\eeqa
for arbitrary  base space background. Since the  base space fields are  $\bar{b}_{\ta\tb}, g_\ta,b_\ta,\prt\vp,\prt\bphi$  and their derivatives, there is no Bianchi identity involved in the base space. Hence the non-gauge-invariant couplings in the base space are all independent. The coefficient of each independent term involves  a specific number which is coming from the DBI contribution, the parameters in \reef{act1} and the parameters $I_1,I_2, \cdots$ of the vector $\tilde{\cI}^\ta$ in the total derivative term \reef{ttot}. The coefficients of all independent terms in the above constraint must be zero. They produce some algebraic equations  that their solution fix the parameters. We have found the following relations for the parameters in \reef{act1}:
\beqa
&&a_2 = -2, a_8= -1 + a_1/2,b_1= 1/24, b_{11}=
 1/6 - a_1/6, b_2 = -1/4, b_4 = 1/4, \nn\\
 &&b_{42} = -2 +
  2 a_1, b_{45 }= -1 + a_1, b_{47} = 2 - a_1, b_5= 1\labell{ab}
\eeqa
and the following vector for the total derivative term:
\beqa
\tilde{\cI}^\ta&=&-2 e^{\varphi} g_\tb \partial^\ta g^\tb -  \partial^\ta\varphi + (2
-  a_1) e^{\varphi} g_\tb \partial^\tb g^\ta + a_1 e^{\varphi} g^\ta
\partial^\tb g_\tb
\eeqa
where the indices are contracted with the base space metric $\bg_{\ta\tb}$. If the spacetime has  no boundary, then the above total derivative terms would be zero. In that case, the parameter $a_1$ would  not be fixed. If one sets $a_1=2$, then the couplings \reef{act1} become exactly the ones have been found in \cite{Corley:2001hg,Garousi:2013gea}. It has been shown in \cite{Corley:2001hg,Garousi:2013gea} that they are fully reproduced by the corresponding disk-level S-matrix elements.

However, if the spacetime $M^{(26)}$ has a boundary  $\prt M^{(26)}$, then the total derivative terms can not be ignored. Then the T-duality of boundary may fix the remaining parameter $a_1$.  If   both spacetime and its boundary have  a circle, \ie $M^{(26)}=S^{(1)}\times M^{(25)}$, $\prt M^{(26)}=S^{(1)}\times \prt M^{(25)}$,  then the D$_{p-1}$-brane which  is transverse to the circle, may end on the boundary,  \ie  the D$_{p-1}$-brane which is along the subspace $M^{(p)}$ in the base space $M^{(25)}=M^{(p)}\times M^{(25-p)}$ may  have the boundary $\prt M^{(p)}$. In that case, the Stokes's theorem in this subspace is
 \beqa
\int_{M^{(p)}} d^{p}\sigma\sqrt{-\bg}\bg^{\ta\tb}\prt_\ta ( e^{-\Phi}{\cal I}_\tb)&=&\int_{\prt {M^{(p)}}}d^{p-1}\tau\,e^{-\Phi}\sqrt{|g|}\,\bg^{\ta\tb}n_{\ta}  {\cal I}_{\tb}\labell{Stokes}
\eeqa
where   $n_\ta=\prt_\ta X^\tmu n_{\tmu}$ and  $n_{\tmu}$ is the normal vector to the boundary $\prt M^{(25)}$ which  is outward-pointing (inward-pointing) if the boundary is spacelike (timelike),  and the boundary in the static gauge is specified by the  functions  $\sigma^\ta=\sigma^\ta(\tau^{\ba})$.  In the square root on the right-hand side, $g$ is determinant of    the induced metric on the boundary, \ie
\beqa
g_{\ba\bb}&=&\frac{\prt \sigma^{\ta}}{\prt\tau^{\ba}}\frac{\prt \sigma^{\tb}}{ \prt\tau^{\bb}}  \bg_{\ta\tb}\labell{bg}
\eeqa
 The coordinates of the boundary $\prt M^{(p)}$ are $\tau^0,\tau^1,\cdots, \tau^{p-2}$.

Using the above  Stokes's theorem, one finds that the contribution of the total derivative terms in the boundary  is
\beqa
\tilde{\cJ}(0)=\frac{\alpha'T_{p-1}}{2}\int_{\prt {M^{(p)}}}d^{p-1}\tau\,e^{-\Phi}\sqrt{|g|}\,n_{\ta} \bigg[2 e^{\varphi} g_\tb \partial^\ta g^\tb + \partial^\ta\varphi - (2
-  a_1) e^{\varphi} g_\tb \partial^\tb g^\ta - a_1 e^{\varphi} g^\ta
\partial^\tb g_\tb\bigg]
\eeqa
The above terms in the boundary indicates that there must be some world-volume couplings on the boundary.

When the spacetime has boundary, the D$_p$-planes in this manifold may end on the boundary. If one writes the spacetime as $M^{(26)}=M^{(p+1)}\times M^{(25-p)}$  where the  D$_p$-plane is along the subspace $M^{(p+1)}$, and  subspace $M^{(p+1)}$ has boundary $\prt M^{(p+1)}$, then  the effective action of D$_p$-plane at  order $\alpha'$ has world-volume couplings \reef{s2} on the bulk of the D$_p$-plane, \ie in  $M^{(p+1)}$, as well as some boundary couplings on the boundary of the D$_p$-plane, \ie in $\prt M^{(p+1)}$. The boundary action should be
\beqa
\prt S_p^1&=&-\frac{\alpha'T_p}{2}\int_{\prt M^{(p+1)}} d^{p}\tau\, e^{-\Phi}\sqrt{|\hg|}\,\sum_{m=0}^{\infty}\prt{\cal L}_m\labell{Genb}
\eeqa
where $\hg$ is the determinant of the pull-back of the pull-back  metric $\tG_{ab}$ on the boundary of D$_p$-plane, \ie
\beqa
\hg_{\ha\hb}&=&\frac{\prt \sigma^{a}}{\prt \tau^{\ha}}\frac{\prt \sigma^{b}}{\prt \tau^{\hb}}\tG_{ab}\labell{hg}
\eeqa
The boundary of D$_p$-plane is specified by  the vectors     $\sigma^{a}(\tau^{\ha})$ where $\tau^0,\tau^1,\cdots \tau^{p-1}$ are coordinates of the boundary,  and $\prt {\cal L}_m$ in \reef{Genb} is the boundary Lagrangian at one-derivative order which includes  various  independent couplings involving the world-volume indies $a, b, \cdots$ and the spacetime  indices $\mu,\nu, \cdots$, evaluated at the boundary of D$_p$-plane. As in the bulk action \reef{s2}, $m$ is the number of pull-back field $\tB_{ab}$ without derivative on it.

 It is implicitly assumed in the T-duality prescription that everything should  be independent of the killing coordinate $y$, hence, to be able to impose the T-duality on the boundary couplings, the boundary should be  specified as $\sigma^a(\tau^{\ha})=(y,\sigma^{\ta}(\tau^{\ba}))$. Then one can show that, in the static gauge and for $X^i=0$,   when D$_p$-plane is along the $y$-direction, the reduction of  $e^{-\Phi}\sqrt{|\hg|}=e^{-\bphi+\varphi/4}\sqrt{|g|}$ where $g$ is determinant of the induced metric  \reef{bg}, and when  D$_{(p-1)}$-plane is orthogonal to  the $y$-direction, the reduction of  $e^{-\Phi}\sqrt{|\hg|}=e^{-\bphi-\varphi/4}\sqrt{|g|}$. The former transforms under the T-duality transformations \reef{buscher} to  the latter. Hence, to find the T-duality constraints on the boundary action \reef{Genb}, one should consider only the T-duality constraint on  $\prt\cL_m$ in \reef{Genb}.

For the boundary couplings at order $m=0$, we consider the following Lagrangian:
\beqa
\prt{\cal L}_0&=&c_1 K_{ab}\tG^{ab}
\eeqa
where $K_{ab}=\prt_a X^\mu\prt_b X^\nu K_{\mu\nu}$ and $K_{\mu\nu}$ is the  extrinsic curvature of  the  spacetime boundary. For the time-like boundary, it is given as
\beqa
K_{\mu\nu}=\nabla_{\mu}n_{\nu}-n_{\mu}n^{\rho}\nabla_{\rho}n_{\mu}
\eeqa
 where   $n_\mu$ is the normal vector to the boundary $\prt M^{(26)}$. It is symmetric and satisfies $n^\mu K_{\mu\nu}=0$. The world-volume reduction of D$_p$-brane and the transverse reduction of D$_{p-1}$ for this boundary couplings are, respectively
 \beqa
 \prt\cL_0^w&=& c_1(\hK_{\ta\tb}\bg^{\ta\tb}+\frac{1}{2}n_{\ta}\prt^\ta\vp)\nn\\
 \prt\cL_0^t&=& c_1(\hK_{\ta\tb}\bg^{\ta\tb}+e^{\vp}n_\ta g_\tb \prt^\ta g^\tb-e^{\vp}n_\tb g_\ta \prt^\ta g^\tb)\labell{rK}
 \eeqa
 where $\hK_{\tmu\tnu}$ is the  extrinsic curvature of  the base space  boundary. We have also removed $g_\ta$ in the terms that have  more than two base space fields, according to  the constraint \reef{cons}. Then one finds that the T-duality constraint for $m=0$ case, \ie
 \beqa
  \prt S_p^{wT}(0)- \prt S_{p-1}^t (0)+\tilde{\cJ}(0)&=&0
  \eeqa
  fixes both the bulk parameter $a_1$ and the boundary parameters $c_1$ to be
  \beqa
  a_1=0&;& c_1=2\labell{ac}
  \eeqa
Hence, the T-duality constraint fixes both the bulk and the boundary actions for  $m=0$ case to be
\beqa
S^1_p+\prt S_p^1&\!\!\!\!\!=\!\!\!\!\!&-\frac{\alpha' T_p}{2}\int d^{p+1}\sigma \, e^{-\Phi}\sqrt { - \tG}\bigg[R ^{ab}{}_{ab}+\frac{1}{6}H_{abc} H^{abc} +\frac{1}{4}H_{ab\mu} H^{ab\mu} -\frac{1}{4}H_{a\mu \nu } H^{a\mu \nu }\labell{L0}\\&&+
\frac{1}{24} H_{\mu \nu \rho } H^{\mu \nu \rho }  -2\Omega_{\mu ab} \Omega^{\mu ab}-\tilde{\nabla}_{a}\tB_{bc} \tilde{\nabla}^{a}\tB^{bc} +
2\nabla_{a}\Phi \nabla^{a}\Phi -2\Omega_{\mu }{}^{a}{}_{a} \nabla^{\mu }\Phi -\nabla_{\mu }\Phi \nabla^{\mu }\Phi\bigg]\nn\\&&
-\frac{\alpha' T_p}{2}\int d^{p}\tau \, e^{-\Phi}\sqrt { | \hg|}\bigg[2K_a{}^a\bigg]\nn
\eeqa
It is interesting to note that the coefficient of the extrinsic curvature is exactly the same as the coefficient of the Hawking-Gibbons term in the spacetime action at the two-derivative level. Note also that if one replaces $\tB\rightarrow \tB+F$, then one finds that the propagators of the massless open string gauge field and the transverse scalar fields receive $\alpha'$-corrections. This is unlike the result in \cite{Corley:2001hg,Garousi:2013gea} in which the study of the boundary couplings did not considered and the standard propagators considered  for these fields.  On the other hand, the coefficients of some of the above couplings are different from those in \cite{Corley:2001hg,Garousi:2013gea}. We will show in the Appendix that the corrections to the propagators, produce  in the S-matrix elements of the leading order DBI action, some contact terms at order $\alpha'$ which change the coefficients of some of the couplings in  the above action such that they become exactly the same  as those in \cite{Corley:2001hg,Garousi:2013gea}.

\subsection {$m=1,2,3,4$ cases}

The T-duality constraint for $m=1,2,3,4$ cases is
\beqa
\sum_{\tm=2}^{5}\Delta\tS^0_1(\tm)+\sum_{\tm=1}^{5}\Delta\tS^1_0(\tm)+\sum_{\tm=1}^{5}\tilde{\cJ}(\tm)&=&0\labell{m1234}
\eeqa
in which the solutions \reef{ab} and \reef{ac} must be  imposed. In this case the   base space background must satisfy the constraint \reef{ta0}. Since the base space is not arbitrary background, unlike the previous case, the above T-duality constraint can not fix all 65 parameters in \reef{act2}. In fact it produces  32 relations. Some of the couplings are fixed and some other are related to each others. We have found the following 32 relations for the parameters in \reef{act2}:
\beqa
&&a_{7 }= -1/2, b_{10 }= -1/16, b_{12 }= 1/2 - a_{6}/4, b_{13 }= -2,
 b_{14 }= 1/16, b_{15 }= 1/4, b_{27 }= -2,\nn\\&& b_{3 }= 1/96, b_{31 }= 2,
 b_{32 }= -1/2, b_{35 }= b_{28}, b_{36 }= -1 - b_{28}, b_{37 }= b_{33},
 b_{38 }= b_{34}, \nn\\&&b_{39 }= -8 b_{25} - b_{33}, b_{40 }= -8 b_{26} - b_{34},
 b_{44 }= -1/2 + 2 a_{4}, b_{46 }= -1/4 + a_{4}, \nn\\&&b_{48 }= -2,
 b_{49 }= 1/2 - a_{4}, b_{50 }= 0, b_{51 }= -2, b_{55 }= -2 b_{24} + b_{41},
 b_{56 }= b_{23} + 16 b_{25}, \nn\\&&b_{57 }= -a_{14} + b_{23}, b_{58 }= b_{24} + 16 b_{26},
  b_{59 }= -b_{23} - 4 b_{24}, b_{6 }= -1/4 + a_{6}/8, \nn\\&&
 b_{60 }= -2 a_{15} + b_{24}, b_{61 }= -b_{23}, b_{8 }= -1/2 + b_{7},
 b_{9 }= 2 - 4 b_{7}\labell{ab345}
\eeqa
and the following vector for the total derivative terms:
\beqa
\tilde{\cI}^\ta&=&-  \frac{1}{4} \bar{b}_{\tb\tc} \bar{b}^{\tb\tc}
\partial^{\ta}\varphi -  b_{25} \bar{b}_{\tb}{}^{\td} \bar{b}^{\tb\tc} \bar{b}_{\tc}{}^{\te}
\bar{b}_{\td\te} \partial^{\ta}\varphi -  b_{26} \bar{b}_{\tb\tc} \bar{b}^{\tb\tc} \bar{b}_{\td\te} \bar{b}^{\td\te}
\partial^{\ta}\varphi \nn\\&&+ \bar{b}^{\ta\tc} \bar{b}_{\tb\tc} \partial^{\tb}\varphi -
\frac{1}{2} b_{23} \bar{b}^{\ta\tc} \bar{b}_{\tb}{}^{\td} \bar{b}_{\tc}{}^{\te} \bar{b}_{\td\te}
\partial^{\tb}\varphi -  \frac{1}{2} b_{24} \bar{b}^{\ta\tc} \bar{b}_{\tb\tc} \bar{b}_{\td\te}
\bar{b}^{\td\te} \partial^{\tb}\varphi +\cdots\labell{tot345}
\eeqa
where dots represent terms that have $g_{\ta}$ without derivative on it. They are zero for the base space background \reef{ta0}. Because of this restriction, the above total derivative derivative terms can also fix only some of the couplings in the boundary action.

The constraint \reef{ta0} on the base space background, causes that some of the couplings in \reef{act2} do not appear in the relations \reef{ab345}.  This is unlike the $m=0$ case in  which the base space background is arbitrary and all parameters in \reef{act1} appear in the relations \reef{ab}. Some of the coefficients of the couplings \reef{act2} which are fixed in \reef{ab345}, \ie $b_{14}$,  $b_{10}$, $b_3$, $b_{15}$, $a_7$, indicate that the couplings in $\cL_2$ which have an extra factor of $\tB_{ab}\tB^{ab}$ with respect to the corresponding couplings in $\cL_0$, satisfy the following relation:
\beqa
\cL_2(b_{14},  b_{10}, b_3, b_{15}, a_7)&=&\frac{1}{4}\tB^{ab}\tB_{ab}\cL_0(b_4,b_2,b_1,b_5,a_2)
\eeqa
The T-duality could not fix $a_4,b_{22},a_{12}$. However, if one chooses these coefficients to be
\beqa
a_4=0,\,b_{22}=-1/24,\, a_{12}=1/4
\eeqa
Then all other couplings  in $\cL_2$ which have an extra factor of $\tB_{ab}\tB^{ab}$ with respect to the corresponding couplings in $\cL_0$, satisfy similar  relation:
\beqa
\cL_2(b_{22}, a_{4}, b_{49}, a_{12}, b_{44},b_{46})&=&\frac{1}{4}\tB^{ab}\tB_{ab}\cL_0(b_{11},a_1,b_{47},a_b,b_{42},b_{45})
\eeqa
The factor $\tB_{ab}\tB^{ab}$ may be resulted from the expansion of the DBI factor, \ie
\beqa
\cL_2(b_{14},  b_{10}, b_3, b_{15}, a_7,b_{22}, a_{4}, b_{49}, a_{12}, b_{44},b_{46})
&=&\sqrt{\det(1+\tG^{ab}\tB_{bc})}\cL_0
\eeqa
The DBI factor should be expanded  and one should keep  $m=2$ terms.

The same relation should be  between the couplings in $\cL_3$ which have an extra factor of $\tB_{ab}\tB^{ab}$ with respect to the corresponding couplings in $\cL_1$, \ie
\beqa
\cL_3(b_{32},  b_{34}, b_{40}, b_{58}, b_{38}, b_{60})&=&\frac{1}{4}\tB^{ab}\tB_{ab}\cL_1(b_{27},b_{28},b_{36},b_{50},b_{35},b_{51})
\eeqa
In the relations \reef{ab345}, $b_{32}=-1/2$ and $b_{27}=-2$. They satisfies the above relation. Also there are  the relations $b_{34}=b_{38}$ and $b_{28}=b_{35}$ which are consistent with the above relation. The factor $\tB_{ab}\tB^{ab}$ may be resulted from the expansion of the DBI factor, \ie
\beqa
\cL_3(b_{32},  b_{34}, b_{40}, b_{58}, b_{38}, b_{60})
&=&\sqrt{\det(1+\tG^{ab}\tB_{bc})}\cL_1\labell{bb1}
\eeqa
The DBI factor should be expanded  and one should keep  $m=2$ terms.

Assuming the DBI factor, the relation between the coupling in $\cL_4$ with coefficients $b_{25}$,  $b_{26}$ and the couplings in $\cL_0$ is
\beqa
\cL_4(b_{25},b_{26})&=&\sqrt{\det(1+\tG^{ab}\tB_{bc})}\cL_0\labell{bb2}
\eeqa
where the DBI factor should be expanded and one should keep  $m=4$ terms. The relations \reef{bb1} and \reef{bb2} are consistent with the T-duality results \reef{ab345} if one chooses the following relations:
\beqa
a_{15}=0,\, b_{24}=-1/2,\, b_{26}=1/32,\, b_{25}=-1/8,\, b_{34}=b_{28}/4,\,b_{38}=b_{35}/4 \labell{aaa}
\eeqa
Hence, it seems the couplings at order $\alpha'$ should have the DBI factor. If one includes the DBI factor in the effective action \reef{s2}, \ie
\beqa
S^1_p&=&-\frac{\alpha' T_p}{2}\int_{M^{(p+1)}} d^{p+1}\sigma \, e^{-\Phi}\sqrt { - \det (\tilde G_{ab}+\tB_{ab})}\,\sum_{m=0}^{\infty}\cL_m\labell{s21}
\eeqa
Then in the Lagrangian $\cL_m$ for $m>1$,  one should removes the independent couplings which have the same structures as those  reproduce by expanding the DBI factor.

To find the boundary couplings for $m=1,2,3,4$ cases, we first note that the constraint \reef{ta0} causes the transverse reduction of $\tB_{\ta\tb},\tG^{\ta\tb}$ to be
\beqa
\tB_{\ta\tb}\,=\,\bar{b}_{\ta\tb}&;&\tG^{\ta\tb}\,=\,\bg^{\ta\tb}
\eeqa
Then one can easily observe that the world-volume reduction of the couplings which involve $\tilde{\nabla}\tB$ produces the base space fields $b_{\ta}$ and $\prt_\tb b_{\ta}$, whereas the transverse reduction of the coupling with structure $n\tB\tilde{\nabla} \tB$ produces $n\bar{b}\prt\bar{b}$.  Hence the boundary action should not have couplings with structures $n\tB\tilde{\nabla} \tB$ and $n\tB\tB\tB\tilde{\nabla} \tB$.

One finds
the transverse reduction of $\tB_{ab}\tB^b{}_c K^{ca}$ is $\bar{b}_{\ta\tb}\bar{b}^\tb{}_\tc \hK^{\tc\ta}$. On the other hand, the world-volume reduction of $\tB_{ab}\tB^b{}_c K^{ca}$ is $\bar{b}_{\ta\tb}\bar{b}^\tb{}_\tc \hK^{\tc\ta}$ plus some other terms that have the vector $b_\ta$ without derivative on it. Under the T-duality \reef{buscher} they are removed by the constraint \reef{ta0}. Hence,  the couplings $\tB_{ab}\tB^b{}_c K^{ca}$ is invariant under the T-duality. Therefore,  the T-duality can not fix the coefficient of the boundary coupling  $\tB_{ab}\tB^b{}_c K^{ca}$.

One can easily observe that for the base space \reef{ta0},  the transverse reduction and the T-duality of the world-volume  reduction of $\Tr(\tB\tB)$ and $\Tr(\tB\tB\tB\tB)$ are $\bar{b}_{\ta\tb}\bar{b}^{\tb\ta}$ and  $\bar{b}_{\ta\tb}\bar{b}^{\tb\tc}\bar{b}_{\tc\td}\bar{b}^{\td\ta}$, respectively. Similarly for any combinations of the tensor $\tB$. Then using the reductions of $K_a{}^a$ in \reef{rK}, and using the reduction for dilaton  in \reef{GB}, one finds that the  terms in the  vector \reef{tot345} are cancelled with the T-duality of the following boundary terms:
\beqa
\prt S_p^1&=&-\frac{\alpha'T_p}{2}\int_{\prt M^{(p+1)}} d^{p}\tau\, e^{-\Phi}\sqrt{|\det(\hg_{\ha\hb})|}\sqrt{\det(1+\tG^{ab}\tB_{bc})}\bigg[2K_a{}^a-2n_a\tB^{ac}\tB_{cb}\nabla^b\Phi\nn\\&&\qquad\qquad\qquad\qquad\qquad\qquad\qquad\qquad-b_{23}n_a\tB^{ac}\tB_{ce}\tB^{ed}\tB_{db}\nabla^b\Phi+\cdots\bigg]
\eeqa
where we have also used the coefficients in \reef{aaa}. Note that $\ha,\hb$ are world-volume indices on the boundary $\prt M^{p+1}$ whereas the indices $a,b,\cdots$ are the world-volume indices in the bulk $M^{p+1}$.  The dots represent the terms at $m>4$ levels in which we are not interested, and also the terms with structures   $\tB_{ab}\tB^b{}_c K^{ca}$ and    $\tB_{ab}\tB^b{}_c \tB^{cd}\tB_{de}K^{ea}$ which can not be fixed by the T-duality of the massless closed string fields. They may be fixed by studying the T-duality of massless open string fields in which we are not interested in this paper.

\section{Discussion}

In this paper we have found the independent world-volume couplings at order $\alpha'$ involving up to five     covariant fields $R$, $H$, $\nabla H$, $\nabla \Phi$, $\nabla\nabla \Phi$,$\tilde{\nabla}\tB$,  $\tilde{\nabla}\tilde{\nabla}\tB$, $\Omega$,  $\tilde{\nabla}\Omega$ and $\tB$.  We have found that there are 76 independent  couplings. The assumption that the effective action of the $D_p$-brane at the critical dimension is background independent is then used to find the parameters of the above independent couplings and to find their corresponding boundary  terms when the spacetime has boundary and the D-brane ends on it. That is, we have considered a particular background which has one circle fibred on a base space. In this background, the effective action should satisfy the T-duality constraint. To impose the T-duality on the closed string fields of these couplings, one has to use the static gauge  to fix the diffeomorphism symmetry. We have shown that in this gauge, the T-duality is satisfied for the  specific base space background \reef{ta0} for the couplings with  more than two fields. The T-duality  constraint  fixes all couplings at the two-field level, and their corresponding boundary term. They are fully consistent with the corresponding disk-level S-matrix elements. Because of the restriction on the base space background, the T-duality is failed to  fix all couplings at more than two-field levels. However, those couplings that are fixed indicate that both the bulk and boundary actions should have the DBI factor.

We have seen that because of the restriction \reef{ta0} on the base space background, we could find only 32 relations \reef{ab345} between the parameters of the independent couplings at orders $m=1,2,3,4$. If one includes the independent couplings at orders $m>4$, we do not expect that the T-duality constraint would fix all parameters at order $m\leq 4$. The T-duality constraint would produce some relations between the independent couplings   at orders $m>4$. However,  some of the unfixed parameters at orders $m\leq 4$ might be fixed by the T-duality on the couplings at orders $m>4$. We expect, if one finds a way to impose the T-duality constraint on the D-brane effective action in covariant form (without using the static gauge), then there would be no constraint on the background because the would-volume component of the second fundamental form \reef{O1} is zero in covariant form without using any restriction on the background. In that case,  the T-duality may fix all couplings at each order of $m$.

The O$_p$-plane effective action has no open string couplings, no couplings that have odd number of transverse indices on metric and dilaton and their corresponding derivatives, and has no couplings that have even number of transverse indices on B-field and its corresponding derivatives \cite{Polchinski:1996na}. This $Z_2$ orientifold projection removes all $m>0$ terms in \reef{s2} and removes $b_{11},a_1,a_2,a_8, b_{42}$ terms in \reef{act1}. It also removes the $\nabla_i\Phi\nabla^i\Phi$ part in the term with coefficient $b_{45}$ and removes the terms with even transverse indices in the couplings with coefficients $b_4,b_2,b_1$. This $Z_2$ projection should be also applied for the reduction of the couplings on the circle. This projects the final constraint on the couplings, \ie \reef{mm}, into two parts. The $Z_2=+1$ corresponds to the O-plane constraint. The DBI contribution \reef{DS0} for $m=0$ case has $A^0_{\ta\tb}=\bg_{\ta\tb}$ and  $ A^1_{\ta\tb}=0$.  Under the $Z_2$ project, $\Delta\vp\rightarrow \Delta\vp_+$.  The T-duality constraint \reef{mm} then fixes  the O$_p$-plane couplings as
\beqa
S_p+\prt S_p&\!\!\!\!\!=\!\!\!\!\!&-\frac{\alpha' T'_p}{2}\int d^{p+1}\sigma \, e^{-\Phi}\sqrt { - \tG}\bigg[R ^{ab}{}_{ab}-\frac{1}{8}H_{abi} H^{abi} +\frac{1}{24}H_{ijk } H^{ijk}+
\nabla_{a}\Phi \nabla^{a}\Phi \bigg]\nn\\&&
-\frac{\alpha' T'_p}{2}\int d^{p}\tau \, e^{-\Phi}\sqrt { | \hg|}\bigg[2K_a{}^a\bigg]\nn
\eeqa
where $T'_p$ is the tension of O$_p$-plane. The above couplings are the $Z_2=+1$ projection of the D-brane couplings in \reef{L0} up to the overall factor. For the couplings at the higher orders of $\alpha'$, we expect, as long as the T-duality fixes the D-brane couplings for $m=0$ case, up to an overall factor, then the $Z_2=+1$ projection on the couplings would  produce the corresponding O-plane couplings up to an overall factor. This overall factor for D-brane and for O-plane  is different. If the T-duality fixes the D-brane couplings up to some parameters, then the O-plane couplings again would be found by the $Z_2=+1$ project. However, the unfixed parameters for D-brane and for O-planes are different numbers that should be fixed by some other methods. Hence, in general, the O-plane couplings are not the $Z_2=+1$ projection of the D-brane couplings up to the  overall tension factor. For example, the Chern-Simons couplings for D-brane and O-plane at order $\alpha'^2$ are the same up to an overall factor, whereas at the higher orders of $\alpha'$ they are not the same up to an overall factor \cite{Morales:1998ux}.

To find the independent world-volume couplings on D-brane, we are allowed to impose the spacetime equations of motion, however, we are not allowed to use the field redefinition for the closed string fields. This causes that the independent couplings include the two-field coupling $(\tilde{\nabla}\tB)^2$. If the spacetime has boundary and the D-brane ends on it, then the T-duality of the boundary term fixes the coefficient of this term to be non-zero. Hence, upon the replacement $\tB\rightarrow \tB+F$, one finds the couplings at order $\alpha'$ change the standard propagator of the  gauge field. The T-duality also fixes the coefficients of the square of the second fundamental form such that the transverse scalar fields also have non-standard propagators. If one considers the D-brane in the presence of the  constant background $B$-field, then one expects that the propagator of the gauge field on this D-brane to be non-standard too. In fact, it is known  that the couplings of two $\tilde{\nabla}F$ and two, four, six, and more $F$'s are non-zero. Upon the replacement $F\rightarrow F+\tB$ and for constant $\tB$, one finds couplings of $\tilde{\nabla}F\tilde{\nabla}F$ and an arbitrary number of constant $B$-field. Hence, the propagator of the gauge field in the presence of constant background is not the standard propagator that is produced by the DBI action. On the other hand, as we have shown in the Appendix, the non-standard propagator produces some contact terms at order $\alpha'$ when one considers the $s$-channel in the S-matrix elements of the DBI action. These couplings should be added to the contact terms that are produced by the standard propagator.

The D-brane world-volume contact terms of two gravitons at order $\alpha'$ in the presence of constant $B$-field background have been found in \cite{Ardalan:2002qt} by using the standard propagators. One the other hand, it has been shown in \cite{Ardalan:2002qt} that the contact terms are not reproduced by  covariant couplings. It would be interesting to include the contact terms resulting from the non-standard propagator to the contact terms found in \cite{Ardalan:2002qt} to see if they can be reproduced by some covariant couplings. If that would be the case, then one must conclude that the parameter $a_1$ in \reef{ab} must be zero even if the spacetime has no boundary. This would then confirm  that the effective action must be background independent, \ie if $a_1=0$  for the background which has boundary, it would remain zero for any other background which may have no boundary.

It has been observed in \cite{Garousi:2022rcv} that the series of the open string gauge field couplings at order $\alpha'$ which are contracted with the inverse of the pull-back metric $\tG_{ab}$ and are consistent with T-duality and S-matrix elements \cite{Karimi:2018vaf}, can not be written in compact form in terms of  the inverse of the tensor $h_{ab}=\tG_{ab}+\tB_{ab}+F_{ab}$. In this paper, however, we have shown that the massless closed string fields in the action should have the DBI factor $\sqrt{-\det(h_{ab})}$.  This observation has been also made  in \cite{Karimi:2018vaf} for the couplings of two fundamental forms and two, four and  six  gauge fields. Hence, we expect, in general, the covariant fields with world-volume indices to be  contracted with the inverse of the pull-back metric, however, the tensor $h_{ab}=\tG_{ab}+\tB_{ab}+F_{ab}$ appears as the overall DBI factor in the actions. Therefore, we expect the couplings of two gravitons in the presence of constant $B$-field found in \cite{Ardalan:2002qt}, after including the contact terms of the non-standard propagator discussed in the pervious paragraph, to have the overall DBI factor and a series  of covariant couplings in terms of the Riemann curvature, the second fundamental form and various powers of $B$-field that are contracted with $\tG^{ab}$.

We have seen that the T-duality at the two-field level fixes the action up to the parameter $a_1$. If there is no boundary it would remain arbitrary. For $a_1=2$, then the propagators of the massless open string fields do not receive $\alpha'$ corrections. However, if there is a  boundary, then the T-duality fixes this to $a_1=0$, which changes the propagators to the non-standard form.  Assuming the effective action of the D-brane in the critical dimension to be  background independent, then one should not consider $a_1$ to be arbitrary because it is fixed in a particular background which has boundary. Hence, the background independence dictates that the effective action at order $\alpha'$ to be the one in \reef{L0}, not the one has been found in \cite{Corley:2001hg,Garousi:2013gea}. We expect the same thing for the propagators of the D-branes in the superstring theory. The curvature squared   terms of the D-brane action at order $\alpha'^2$ have been found in \cite{Bachas:1999um} by considering the low energy contact terms of the disk-level S-matrix element of two gravitons \cite{Garousi:1996ad} at order $\alpha'^2$. In this study, the massless poles of the amplitude at order $\alpha'^0$ are assumed to be reproduced by the DBI action which produces the standard propagators for the gauge field and for the transverse scalar fields. However, if the propagators receive $\alpha'^2$-corrections, then the contact terms at order $\alpha'^2$ would  be changed and the curvature squared terms would also be changed. In fact,  the gauge field couplings at order $\alpha'^2$ that have been found in \cite{Wyllard:2000qe} by the the boundary state method, do have term $(\tilde{\nabla}\tilde{\nabla}F)^2$. However, it has been argued in \cite{Wyllard:2000qe} that this term can be  removed by the appropriate field redefinitions. On the other hand, the $B$-field gauge symmetry requires that the world-volume couplings  have either the $B$-field strength $H$ or the combination $\tB+F$. Since there is no field redefinition on the D-brane for the $\tB$-field, the $B$-field gauge symmetry requires to have no field redefinition for the open string gauge field $F$ either.  To check if the propagators receive $\alpha'^2$-corrections, one may extend the calculations in this paper to the superstring theory on a spacetime manifold which has boundary. If the coefficient of the world-volume  coupling $(\tilde{\nabla}\tilde{\nabla}\tB)^2$ is non-zero, then the replacement $\tB\rightarrow \tB+F$ would produce the gauge field coupling $(\tilde{\nabla}\tilde{\nabla}F)^2$ which changes the propagator. It would be interesting to perform this calculation at $m=0$ level, to find NS-NS couplings at order $\alpha'^2$ by the T-duality and  to see if the propagators on the world-volume of D-branes are standard or not. Similar calculation for O-plane which has no open string fields,  has been done in \cite{Robbins:2014ara,Garousi:2014oya,Akou:2020mxx,Mashhadi:2020mzf}.
\\

{\bf Acknowledgements}:  This work is   supported by Ferdowsi University of Mashhad under grant  3/54460(1400/02/22).
\\
\\
\\

{\bf Appendix}
\\

In this appendix we are going to show that the couplings in \reef{L0} are fully consistent with the S-matrix elements. It has been shown in \cite{Corley:2001hg,Garousi:2013gea} that the couplings \reef{act1} with the coefficients \reef{ab} are fully consistent with the S-matrix element for $a_1=2$. To show that for $a_1=0$ case also the couplings are consistent with the S-matrix elements, we consider the terms in \reef{act1} which depend on $a_1$ and also $\Omega_{\mu ab} \Omega^{\mu ab}$. They are
\beqa
\cL_0&=&(\frac{1}{6}-\frac{a_1}{6})H_{abc} H^{abc}
  +  a_{1} \Omega_{\mu }{}^{b}{}_{b} \Omega^{\mu a}{}_{a} - 2 \Omega_{\mu ab} \Omega^{\mu ab} +  (-1+\frac{a_1}{2})\tilde{\nabla}_{a}(\tB_{bc}+F_{bc}) \tilde{\nabla}^{a}(\tB^{bc} +F^{bc})\nn\\&&+
(2-a_1)\nabla_{a}\Phi \nabla^{a}\Phi -(2-a_1)\Omega_{\mu }{}^{a}{}_{a} \nabla^{\mu }\Phi -(1-a_1)\nabla_{\mu }\Phi \nabla^{\mu }\Phi \labell{act11}
\eeqa
we have also replaced $\tB$ by $\tB+F$. For $a_1=2$, one finds the propagators for the open string gauge field and the transverse scalar fields receive no $\alpha'$-correction. The propagators are produced only by the couplings in the DBI action, \ie
\beqa
S_p^0&=&-T_p\int d^{p+1}\sigma\bigg[1+\frac{1}{2}\prt_a X^i\prt^a X^j\eta_{ij}+\frac{1}{4}F_{ab}F^{ab}+\cdots\bigg]
\eeqa
The propagators are
\beqa
(G^{XX})^{ij}\,=\, \frac{i\eta^{ij}}{T_p s}&;&(G^{AA})^{ab}\,=\,\frac{i\eta^{ab}}{T_p s}\labell{G0}
\eeqa
where $s=-k_ak^a$ and $k_a$ is the momentum of the open string field.

For $a_1=0$, the couplings \reef{act11} contain the following two-field terms:
\beqa
S_p^1&=&-\frac{\alpha'T_p}{2}\int d^{p+1}\sigma\bigg[-2\prt_a \prt_b X^i\prt^a\prt^b X^j\eta_{ij}-\prt_cF_{ab}\prt^cF^{ab}+\cdots\bigg]
\eeqa
They change the propagators in \reef{G0} to
\beqa
(G^{XX})^{ij}\,=\, \frac{i\eta^{ij}}{T_p s(1-\alpha' s)}&;&(G^{AA})^{ab}\,=\,\frac{i\eta^{ab}}{T_p s(1-\alpha' s)}\labell{G1}
\eeqa
If one expand them, one finds the propagators in the $s$-channel change to the standard propagators plus some contact terms at higher orders of $\alpha'$, \ie
\beqa
(G^{XX})^{ij}\,=\, \frac{i\eta^{ij}}{T_p s}+\frac{i\alpha' \eta^{ij}}{T_p}+\cdots&;&(G^{AA})^{ab}\,=\,\frac{i\eta^{ab}}{T_p s}+\frac{i\alpha' \eta^{ab}}{T_p}+\cdots\labell{G1}
\eeqa
where dots represent terms at higher orders of $\alpha'$.

Using the DBI action, one can calculate the scattering amplitudes of two massless closed strings in the $s$-channel. They  produce the massless poles at the leading order of $\alpha'$ in the $s$-channel, and some some contact terms at order $\alpha'$. Let us consider the scattering amplitude of two dilatons from D-brane. Using the fact that the closed string fields in the D-brane action are function of the transverse scalar fields and should be Taylor expanded, one finds DBI action produces the following vertex for one on-shell dilaton and one off-shell $X^i$:
\beqa
V(\Phi,X)_i&=&T_p\prt_i\Phi
\eeqa
Using the propagator of the transverse scalar field in \reef{G1}, one finds the $s$-channel produces the following contact terms at order $\alpha'$:
\beqa
C^{\Phi\Phi}&=&-\alpha'T_p\prt_i\Phi\prt^i\Phi=-\frac{\alpha'T_p}{2}(2\prt_\mu\Phi\prt^\mu\Phi-2\prt_a\Phi\prt^a\Phi)
\eeqa
If one adds the above dilaton couplings to the couplings in \reef{act11} for $a_1=0$, one would find the dilaton couplings in \reef{act11} for $a_1=2$.

The  DBI action produces the following vertex for one on-shell $\tB$ and one off-shell $A^a$:
\beqa
V(\tB,A)_a&=&T_p\prt^c\tB_{ca}
\eeqa
Using the propagator of the gauge  field in \reef{G1}, one finds the $s$-channel produces the following contact terms at order $\alpha'$:
\beqa
C^{BB}&=&-\alpha'T_p\prt_c\tB^{ca}\prt^d\tB_{da}=-\frac{\alpha'T_p}{2}(-\frac{1}{3}H^{dca}H_{dca}+\prt^c\tB^{ad}\prt_{c}\tB_{ad})
\eeqa
where we have also used integration by part. If one adds the above $B$-field  couplings to the couplings in \reef{act11} for $a_1=0$, one would find the $B$-field couplings in \reef{act11} for $a_1=2$. Similar calculations can be done for the other couplings as well. Note that the S-matrix of one $\tB$ and one $F$ in \reef{act11} is also zero. Hence, the S-matrix elements for $a_1=0$ are the same as the S-matrix elements for $a_1=2$. On the other hand, it has been shown in \cite{Corley:2001hg,Garousi:2013gea} that the field theory S-matrix element for $a_1=2$ are consistent with the low energy expansion of the disk-level S-matrix element in the string theory. Hence the couplings in \reef{L0} are fully consistent with the corresponding S-matrix elements in the bosonic string theory.


\end{document}